\documentclass{egpubl}
\usepackage[T1]{fontenc}
\usepackage{dfadobe}  

\usepackage{cite}  % comment out for biblatex with backend=biber
\BibtexOrBiblatex
\electronicVersion
\PrintedOrElectronic
\ifpdf \usepackage[pdftex]{graphicx} \pdfcompresslevel=9
\else \usepackage[dvips]{graphicx} \fi

\usepackage{egweblnk} 
\usepackage{amsmath}
\usepackage{amssymb}
\usepackage[caption=false]{subfig}
\usepackage{tikz}
\usepackage[ruled,linesnumbered]{algorithm2e}
\usepackage{qcircuit}
\graphicspath{{fig/}}
\usetikzlibrary{quotes, angles}

\newcommand{\bra}[1]{\left\langle#1\right|}
\newcommand{\ket}[1]{\left|#1\right\rangle}
\newcommand{\etal}{et al.}

\usepackage{color}

\title{Improved Quantum Supersampling for Quantum Ray Tracing}

\author[Xi Lu \& Hongwei Lin]{\parbox{\textwidth}{\centering Xi Lu$^{1,2}$% \orcid{0000-0003-4121-3419}
  and Hongwei Lin$^{1,2}$% \orcid{0000-0002-9337-9624}
  } \\
  {\parbox{\textwidth}{\centering $^1$School of Mathematical Science, Zhejiang University, Hangzhou, 310027, China\\
    $^2$State Key Lab. of CAD\&CG, Zhejiang University, Hangzhou, 310058, China
    }
  }
}

\begin{document}

\maketitle

\begin{abstract}
  Ray tracing algorithm is a category of rendering algorithms that calculate the color of pixels by simulating the physical movements of a huge amount of rays and calculating their energies, which can be implemented in parallel.
  Meanwhile, the superposition and entanglement property make quantum computing a natural fit for parallel tasks.
  Here comes an interesting question, is the inherently parallel quantum computing able to speed up the inherently parallel ray tracing algorithm?
  The ray tracing problem can be regarded as a high-dimensional numerical integration problem.
  Suppose $N$ queries are used, classical Monte Carlo approaches has an error convergence of $O(1/\sqrt{N})$, while the quantum supersampling algorithm can achieve an error convergence of approximately $O(1/N)$.
  However, the outputs of the origin form of quantum supersampling obeys a probability distribution that has a long tail, which shows up as many detached abnormal noisy dots on images.
  In this paper, we improve quantum supersampling by replacing the QFT-based phase estimation in quantum supersampling with a robust quantum counting scheme, the QFT-based adaptive Bayesian phase estimation.
  We quantitatively study and compare the performances of different quantum counting schemes.
  Finally, we do simulation experiments to show that the quantum ray tracing with improved quantum supersampling does perform better than classical path tracing algorithm as well as the original form of quantum supersampling.
  

  
\end{abstract}

\section{Introduction}
\label{sec:introduction}

Ray tracing~\cite{Whitted1980, Cook1984, Kajiya1986, haines2019ray} is a general term of rendering algorithms that calculate pixel colors by simulating all the physical interactions between light rays and the scene.
Since a single ray scatters towards many directions when interacting with an object, and each of the scattered rays scatters towards more directions when interacting with other objects, the total number of rays grow exponentially in the number of interactions.
A standard solution to this problem is Monte Carlo ray tracing, or \textit{path tracing}~\cite{Kajiya1986}, which randomly shoots only one ray at each bounce.
To render an image with high quality, path tracing algorithm require many rays to reduce noise.
In many situations people have to make a trade-off between time cost and quality.
For real-time ray tracing applications where the rendering time is strictly limited, the state-of-the-art GPU can only handle sampling a small amount of rays per pixel, and the quality of result relies heavily on subsequent denoising procedures~\cite{real_time_rendering, Zheng2021Temporally, schied2017spatiotemporal}.
Therefore, a major problem in ray tracing is to reduce the time cost while maintaining the quality.

Quantum computing~\cite{qcqi} is an emerging subject that studies how to perform computational tasks in quantum mechanical systems.
By leveraging the superposition and entanglement of quantum computing, quantum computing has inherent advantages on parallel computational tasks.
As a result, quantum computing shows its computational power by providing spectacular speedup over classical computing in some problems~\cite{Grover1997, Shor1997}.

The idea of introducing quantum computing into computer graphics was early proposed in 2005~\cite{Lanzagorta2005}, which raises many concepts such as quantum Z-buffer, ray tracing and radiosity algorithm.
Later, Caraiman~\cite{caraiman2012quantum} introduced quantum solutions for the polygon visibility and global illumination problems and developed the appropriate quantum algorithms.
In both articles, quantum speedup comes from the superposition of all scene primitives.
Johnson~\cite{Eric2016} proposed quantum supersampling as the quantum variant of Monte Carlo integration in ray tracing, and did simulation experiments on binary image filtering to show that quantum supersampling can reduce mean pixel error faster.
Shimada \etal~\cite{Shimada2020} also did experiments on binary image filtering using quantum coin (QCoin) method originally proposed in~\cite{Abrams1999}.
Recently Santos\etal~\cite{Santos2022} investigated on using quantum computing for intersection searching subroutine in ray tracing, which provides a quadratic complexity in scene complexity.

\begin{figure*}
    \centering
    \subfloat[Classical path tracing.]{
        \includegraphics[width=.4\linewidth]{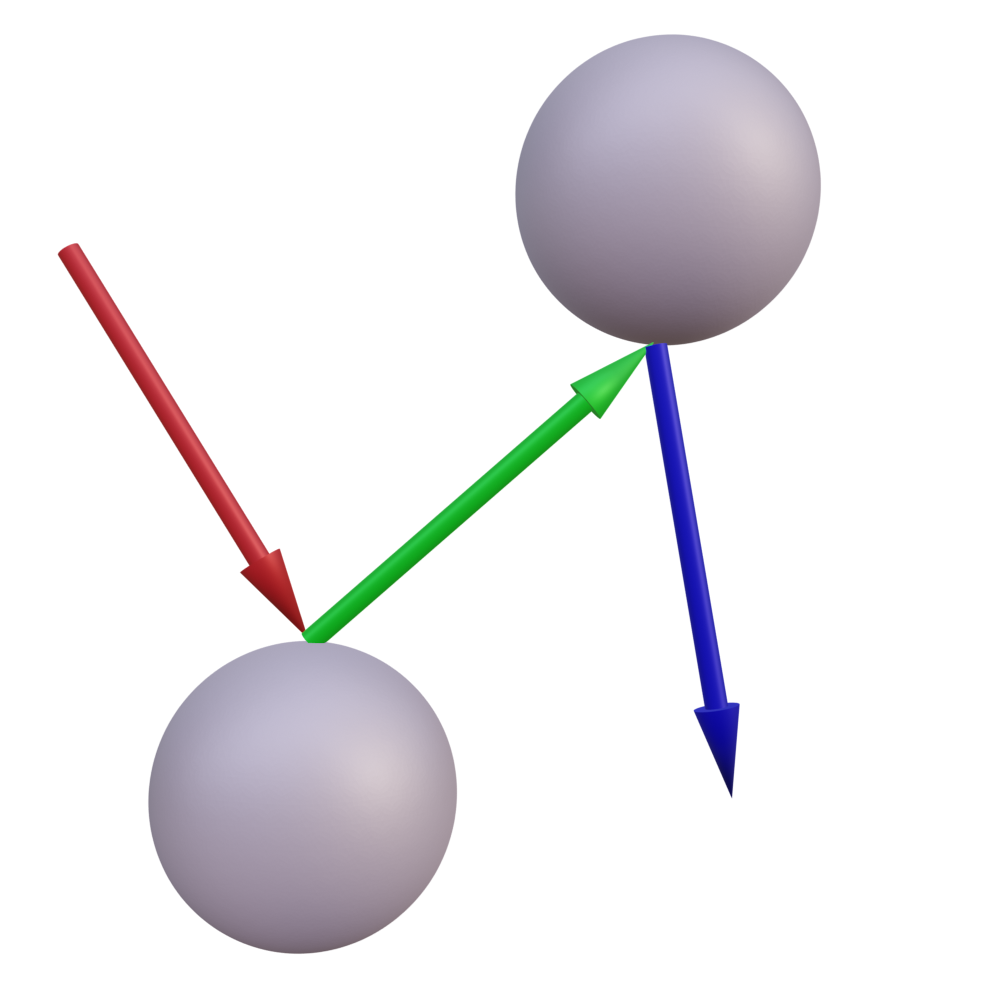}
        \label{fig:classical-ray-tracing}
    }
    \qquad
    \subfloat[Quantum ray tracing.]{
        \includegraphics[width=.4\linewidth]{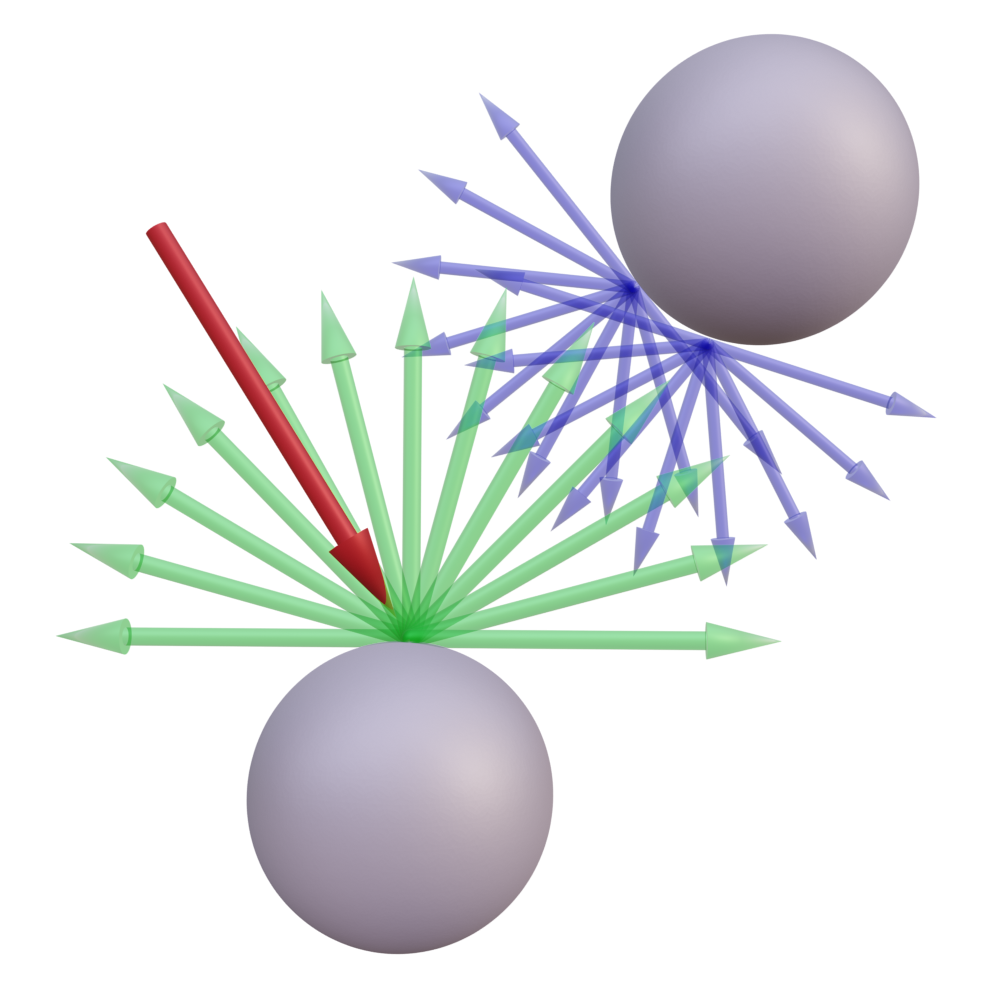}
        \label{fig:quantum-ray-tracing}
    }
    \caption{Classical path tracing only traces one ray at a time, while quantum ray tracing can trace numerous rays as a superposition in one shot.}
    \label{fig:ray-tracing}
\end{figure*}

There are generally two potential approaches for quantum speedup in ray tracing algorithm.
One is to store the scene primitives in a superposition and apply quantum minimum search\cite{Durr1996} to find the nearest intersection between each ray and the scene, to obtain a quadratic speedup, as well studied in~\cite{Santos2022}.
% This approach requires to frequently read the scene primitives from classical memory and then superpose them in quantum memory, and the superposed state is not clonable.
% Also, we do not know yet whether this operation can be implemented in sublinear time.
The other is to store the ray paths in a superposition, as illustrated in Fig.~\ref{fig:ray-tracing}, such that an astronomical amount of rays, at the cost of a logarithmic space and time, can be traced simultaneously to reduce error.
Plus, we do not have the same worry as the first approach, since the rays are procedurally generated instead of reading from classical memory.
This approach is used in quantum supersampling and quantum coin method.

As is shown in~\cite{Eric2016}, the image produced by quantum supersampling (QSS) contains many detached noisy dots. In this paper, we improve quantum supersampling for quantum ray tracing, by replacing the standard quantum Fourier transformation (QFT) based phase estimation algorithm (QFT-PEA) with more robust quantum counting schemes.
The quality of the output image relies dominantly on the quantum counting scheme.
For example, the quantum counting scheme used in QSS is the standard QFT-PEA, which outputs a random variable that is concentrated in a vicinity of the ground truth, but has a long tail in the meantime.
We quantitatively analyze the performances of different quantum counting schemes.
We also propose the QFT-based adaptive Bayesian phase estimation (QFT-ABPEA) as a substitute for QFT-PEA in QSS.

Finally, we build a 3D scene, use a large enough number of samples with the \textit{Blender cycles} renderer to generate the ground truth, and simulate the quantum noise by sampling random numbers from the theoretical distribution, to simulate the quantum ray tracing result.
We also use a limited number of samples with Blender cycles as a representation of classical path tracing result, and show that quantum ray tracing does perform better than classical ray tracing, conditioned on similar computational cost.

Our contributions are as follows.
\begin{itemize}
    \item Analyze the performances of some existing quantum counting schemes that are potential substitutes of QFT-based phase estimation algorithm in QSS.
    \item Propose a QFT-based adaptive Bayesian phase estimation algorithm.
    \item Apply improved QSS to ray tracing algorithm. Do experiments that simulate the workflow of ray tracing, and simulate the images rendered by classical and quantum ray tracing to prove that quantum ray tracing does have better visual performance over classical ray tracing, if appropriate quantum counting scheme is used.
\end{itemize}

The structure of the paper is as follows.
In Section~\ref{sec:preliminary} we present basic knowledge about quantum computing, including fundamental concepts, quantum phase estimation and quantum counting.
In Section~\ref{sec:algorithm}, we present a framework for quantum ray tracing, analyze the performances of several schemes and propose a QFT-based adaptive Bayesian phase estimation scheme.
Then in Section~\ref{sec:experiment}, we do some simulation experiments on real images rendered by classical ray tracing, QCoin, and QSS with different schemes, to show that improved QSS does have a better visual performance than classical ray tracing, the original form of QSS and QCoin method, conditioned on the same computational cost.
In Section~\ref{sec:discussion} and Section~\ref{sec:conclusion}, we make some discussions and conclusions.

\section{Preliminary}
\label{sec:preliminary}

\subsection{Fundamental concepts of Quantum Computing}

All stories began in the 1980s when Feynman suggested that quantum mechanics might be more computationally powerful than Turing machine in some problems like simulating the physical world~\cite{Feynman1982,Feynman1986}.
By substituting classical bits for quantum bits, or {\itshape qubits}, which can be not only in the states $\ket{0}$ and $\ket{1}$ but also their superposition $a\ket{0}+b\ket{1}$ where $a,b\in\mathbb{C}$ and $|a|^2+|b|^2=1$, quantum computing obtains many interesting features like entanglement, reversibility, parallelism, no-cloning and non-orthogonal indistinguishability~\cite{qcqi}.

Quantum computing follows the gate model with three general steps: initializing all qubits into zero state, performing unitary transformations, and finally measuring them to turn quantum information into classical one.
Operations on qubits are implemented by {\itshape quantum gates}, the quantum variants of classical logic gates.
As we know, any classical computing circuit can be constructed with NOT gates, AND gates and COPY gates.
Thus, a quantum computer can simulate a classical computer, by restricting the qubit states to $\{\ket{0},\ket{1}\}$, and using the Toffoli gate, X gate and CNOT gate to replace the AND gate, NOT gate and the COPY gate in classical computers, respectively.
\begin{equation*}
    \begin{aligned}
        \text{Toffoli: }& \ket{a}\ket{b}\ket{0} &\mapsto& \ket{a}\ket{b}\ket{a\text{ and }b} \\
        \text{X: }& \ket{a} &\mapsto& \ket{\text{not }a} \\
        \text{CNOT: }& \ket{a}\ket{0} &\mapsto& \ket{a}\ket{a}
    \end{aligned}
\end{equation*}

Furthermore, due to the reversibility of the three gates above, the quantum implementation of a classical function $j\mapsto f(j)$ should be of the following form,
\begin{equation}
    \ket{j}\ket{0} \mapsto \ket{j}\ket{f(j)},
\end{equation}
which we abbreviate as $\ket{j}\mapsto\ket{j}\ket{f(j)}$ throughout this paper.
Here $\ket{j}$ and $\ket{f(j)}$ are quantum registers that use several qubits to store various data structures like integers and real numbers.
% If we avoid using any measurement gate in the circuit of computing $f$, then we can make full use of the linearity and reversibility of unitary gates.
It follows immediately that when a superposition state $\sum_jx_j\ket{j}$ is inputted, where $x_j$s are arbitrary complex coefficients, the same quantum circuit performs the following linear transformation,
\begin{equation}
    \sum_jx_j\ket{j} \mapsto \sum_jx_j\ket{j}\ket{f(j)},
    \label{eq:oracle-f}
\end{equation}
due to the linear property.
We call such circuits {\itshape quantum linear circuits}.
Any classical circuit can be turned into quantum linear circuit in the brute-force way above, though there are works studying more efficient approaches, like quantum circuit for addition and comparison~\cite{QuantumAddition}.

Though several evaluations of the function $f$ are computed in one query, we cannot read them all out directly.
Once we measure the state on the computational basis and get access to a specific $f(j_0)$, the whole state must collapse to the basis state $\ket{j_0}\ket{f(j_0)}$, and the information of other evaluations is lost forever.
Anyway, we have to design clever algorithms to make the best use of quantum parallelism.
Some of such examples are Grover's search~\cite{Grover1997}, minimum finding~\cite{Durr1996}, quantum counting~\cite{QuantumCounting1998}, and quantum numerical integrals~\cite{Abrams1999}.

\subsection{Quantum Phase Estimation}

Given a quantum circuit that performs unitary transformation $U$, and an eigenstate $\ket{\psi}$ of $U$ such that 
\begin{equation}
    U\ket{\psi}=e^{2\pi i\varphi}\ket{\psi},
\end{equation}
the \textit{phase estimation algorithm} (PEA)~\cite{Kitaev1995,Shor1997} provides an efficient way to estimate $\varphi$.
For general state $\ket{\psi}$ that is not necessarily an eigenstate, let 
\begin{equation}
    \ket{\psi} = \sum_jc_j\ket{\psi_j},
\end{equation}
be the orthogonal decomposition onto the eigenspaces of $U$, where $\ket{\psi_j}$is an eigenstate of $U$ with respect to eigenvalue $e^{2\pi i\varphi_j}$, then PEA returns an estimation of $\varphi_j$ with probability $|c_j|^2$.
The PEA is considered the source of quantum speedup of the celebrated Shor's integer factorization algorithm~\cite{Shor1997}.

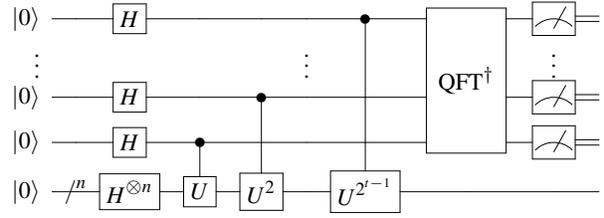
\begin{figure}[ht]
    $$\Qcircuit @C=1.0em @R=0.5em {
        \lstick{\ket{0}} & \qw & \gate{H} & \qw & \qw & \qw & \ctrl{4} & \multigate{3}{\mathrm{QFT}^\dagger} & \meter & \cw \\
        \lstick{\vdots} & & & & & \vdots & & \nghost{QFT^\dagger} & \vdots & & \\
        \lstick{\ket{0}} & \qw & \gate{H} & \qw & \ctrl{2} & \qw & \qw & \ghost{\mathrm{QFT}^\dagger} & \meter & \cw \\
        \lstick{\ket{0}} & \qw & \gate{H} & \ctrl{1} & \qw & \qw & \qw & \ghost{\mathrm{QFT}^\dagger} & \meter & \cw \\
        \lstick{\ket{0}} & \qw{/^n} & \gate{H^{\otimes n}} & \gate{U}& \gate{U^2} & \qw & \gate{U^{2^{t-1}}} & \qw & \qw & \qw
    }$$
    \caption{ The quantum circuit of QFT-PEA. }
    \label{fig:PEA}
\end{figure}

The well-known \textit{Quantum Fourier Transformation}(QFT)-based form of PEA uses the circuit shown in Fig.~\ref{fig:PEA}, where $H^{\otimes t}$ gate means applying Hadamard gate to each of the $t$ qubits, $QFT^\dagger$ is the inversed QFT circuit.
The number of queries to $U$ is $T-1$.

From the textbook~\cite{qcqi} we can find that the output $\tilde{\varphi}$ of PEA obeys the following distribution,
\begin{equation}
    P(\tilde{\varphi}|\varphi) = \left(\frac{\sin(T\pi(\tilde{\varphi}-\varphi))}{T\sin(\pi(\tilde{\varphi}-\varphi))}\right)^2,
    \label{eq:dist-PEA}
\end{equation}
where $T=2^t$ and $\tilde{\varphi}\in\{0,1/T,2/T,\cdots,1\}$.
From Eq.~\label{eq:dist-PEA} we know that QFT-PEA is accurate when $\varphi$ is an integer multiplication of $T^{-1}$, and shows the biggest noise when $\varphi$ is a half integer multiplication of $T^{-1}$.
The probability of estimating within accuracy $1/T$ is at least $8/\pi^2$.

% Anyway, current hardware development allows only a limited amount of quantum memory and gate operations before decoherence.
% Therefore, compared to pure quantum algorithms that run entirely on quantum realm, the hybrid quantum-classical algorithms that run with less quantum memory and time in one shot and relies on post-processing on classical computers are more likely to be implemented on NISQ (Noise Intermediate-Scale Quantum computers) in the near future.

The phase estimation problems are sometimes called in a more general name, the amplitude estimation problem.
That is, given a unitary transformation $U:\ket{0}\rightarrow\sqrt{1-a^2}\ket{0}+a\ket{1}$, estimate $a$.
The schemes for amplitude estimation are still a research hotspot in recent years~\cite{Svore2013,Wiebe2016,Wie2019,Suzuki2019,Grinko2021}.

\subsection{Quantum Counting Scheme}

Given any Boolean function $f:\{0,1,\cdots,N-1\}\rightarrow\{0,1\}$, where $N=2^n(n\in\mathbb{Z}_+)$, as well as the quantum circuit that performs the transformation,
\begin{equation}
    O_f: \ket{j} \mapsto \ket{j}\ket{f(j)}, \quad(j=0,1,\cdots,N-1)
\end{equation}
the quantum counting algorithm~\cite{QuantumCounting1998} can estimate the sum $S=\sum_{j=0}^{N-1}f(j)$ with a quadratic faster convergence rate over classical Monte Carlo counting.

The key idea of quantum counting is to construct a unitary transformation whose eigenvalue contains information about $S$.
The celebrated Grover's iteration~\cite{Grover1997} is exactly such a unitary transformation.
Let
\begin{eqnarray}
    \ket{\alpha} & = & \frac{1}{\sqrt{N-S}}\sum_{f(j)=0}\ket{j},\\
    \ket{\beta} & = & \frac{1}{\sqrt{S}}\sum_{f(j)=1}\ket{j},\\
    \ket{u} & = & \frac{1}{\sqrt{N}}\sum_{j=0}^{N-1}\ket{j}.% = \cos(\pi\varphi)\ket{\alpha} + \sin(\pi\varphi)\ket{\beta}.
\end{eqnarray}

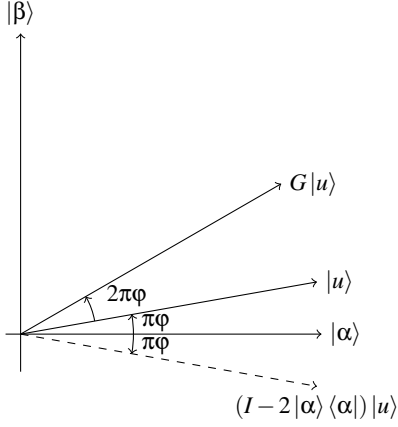
\begin{figure}
    \centering
    \begin{tikzpicture}
        \coordinate (O) at (0, 0);
        \coordinate (X) at (4, 0);
        \coordinate (Y) at (0, 4);
        \coordinate (U) at (10: 4);
        \coordinate (G) at (30: 4);
        \coordinate (R) at (-10: 4);
        \draw[->] (-0.2,0) -- (X) node[right] {$\ket{\alpha}$};
        \draw[->] (0,-0.5) -- (Y) node[above] {$\ket{\beta}$};
        \draw[->] (O) -- (U) node[right] {$\ket{u}$};
        \draw[->] (O) -- (30: 4) node[right] {$G\ket{u}$};
        \draw[->, dashed] (O) -- (-10: 4) node[below] {$(I-2\ket{\alpha}\bra{\alpha})\ket{u}$};
        \pic["$\pi\varphi$", draw=black, ->, angle eccentricity=1.2, angle radius=1.5cm]{angle=X--O--U}; 
        \pic["$\pi\varphi$", draw=black, <-, angle eccentricity=1.2, angle radius=1.5cm]{angle=R--O--X}; 
        \pic["$2\pi\varphi$", draw=black, ->, angle eccentricity=1.5, angle radius=1.0cm]{angle=U--O--G}; 
    \end{tikzpicture}
    \caption{An illustration of Grover's iteration.}
    \label{fig:Grover-search}
\end{figure}

As illustrated in Fig.~\ref{fig:Grover-search}, a Grover's iteration~\cite{Grover1997} $G$ consists of two reflections, one about $\ket{\alpha}$ and the other about $\ket{u}$, that is,
\begin{equation}
    \begin{aligned}
        G = & \left(I-2\ket{u}\bra{u}\right) \left(I-2\ket{\alpha}\bra{\alpha}\right)
        \\ = & H^{\otimes n}O_0H^{\otimes n}P_f,
    \end{aligned}
\end{equation}
where
\begin{equation}
    O_0=I-2\ket{0}\bra{0},
\end{equation}
and $H^{\otimes n}$ is to apply Hadamard transformation to each of the $n$ qubits, and $P_f$ is called the phase oracle of $O_f$ which performs the transformation,
\begin{equation}
    P_f: \sum_j x_j \ket{j} \mapsto \sum_j (-1)^{f(j)} x_j \ket{j}.
\end{equation}

Therefore, $U$ acts as a rotation by twice the angle between $\ket{u}$ and $\ket{\alpha}$ on the plane spanned by $\{\ket{\alpha},\ket{\beta}\}$.
When restricted in this plane, the eigenvalues of such a plane rotation is $e^{\pm2\pi i\varphi}(0\le\varphi<1/2)$, where
\begin{equation}
    \sin(\pi\varphi) = \sqrt{\frac{S}{N}}, 
    \label{eq:varphi-S}
\end{equation}
is the rotation angle, and the corresponding eigenstates are
\begin{equation}
    \ket{\psi_{\pm}} = \frac{1}{\sqrt{2}}(\ket{\alpha}\mp i\ket{\beta}).
\end{equation}

Since
\begin{equation}
    \ket{u} = e^{i\pi\varphi}\ket{\psi_+}+e^{-i\pi\varphi}\ket{\psi_-},
    \label{eq:uniform-superposition}
\end{equation}
we can defer that when applying PEA to $U$ with quantum state $\ket{u}$, we can get the output $\varphi$ or $1-\varphi$ with equal probability.
We can easily distinguish them, as $0\le\varphi<\frac{1}{2}$.
From Eq.~\eqref{eq:dist-PEA} we see the quantum counting result obeys the random distribution,
\begin{equation}
    P(\tilde{\varphi}|\varphi) = \begin{cases}
        \left(\frac{\sin(T\pi(\tilde{\varphi}-\varphi))}{T\sin(\pi(\tilde{\varphi}-\varphi))}\right)^2+&\left(\frac{\sin(T\pi(\tilde{\varphi}+\varphi))}{T\sin(\pi(\tilde{\varphi}+\varphi))}\right)^2,\\&\tilde{\varphi}=\frac{1}{T},\frac{2}{T},\cdots,\frac{1}{2}-\frac{1}{T};
        \\\left(\frac{\sin(T\pi(\tilde{\varphi}-\varphi))}{T\sin(\pi(\tilde{\varphi}-\varphi))}\right)^2,&\tilde{\varphi}=0,\frac{1}{2}.
    \end{cases}
    \label{eq:dist-QC}
\end{equation}

Another approach is the \textit{Amplitude Amplification} (AA), which uses Grover's iteration as amplitude amplification, and the information about amplitude is extracted via repeated direct measurements.
For example, by applying $O_f$ directly to the uniform superposition state $\ket{u}$ we obtain $\frac{1}{\sqrt{N}}\sum_j\ket{j}\ket{f(j)}$, so the measurement on the second register will output 1 with probability $\sin^2(\pi\varphi)=S/N$.
If $M$ times of Grover's iteration is applied to $\ket{u}$ before measurement, the probability becomes $\sin^2((2M+1)\pi\varphi)$.
With well-designed schemes like \cite{Abrams1999,Suzuki2019,Aaronson2020} the information about $\varphi$ can also be extracted with high precision.

\section{Algorithm}
\label{sec:algorithm}

When dealing with a numerical integration problem, classical Monte Carlo method randomly evaluates $N$ samples to get an estimation with an error convergence of $1/\sqrt{N}$.
In quantum computers, a large range of samples can be made into a superposition state, and thus can be computed in one shot.
Based on this idea, Johnson~\cite{Eric2016} proposed quantum supersampling (QSS), and experimentally proved it to have a faster convergence than Monte Carlo method.
However, their result images contain many detached noisy dots that severely affect the quality of image, as they use a non-robust QFT-based phase estimation (QFT-PEA) as the quantum counting scheme.
In this section, we present a framework of quantum ray tracing, then propose improved QSS by replacing the QFT-based phase estimation with more robust quantum counting schemes.

\subsection{Framework of quantum ray tracing}

In classical path tracing, many ray paths are required to calculate the color of a single pixel as accurately as possible.
In quantum computing, the information of all those rays can be stored in a superposition.
All we need is an extra register that stores the ID of each superposed ray, in the form $\sum_{ID}\ket{ID}\ket{ray_{ID}}$.
Moreover, the quantum memory used for storing those IDs uses only a logarithmic space, while the memory used for storing path information like origins and directions is shared in superposition, so the biggest advantage of quantum ray tracing is that we can reduce the sampling error to a negligible level by using an astronomical number of rays.

As is introduced, we can assume that we are given a ray tracing oracle implementing the following transformation,
\begin{equation}
    O_f(pixel, channel): \sum_{j=0}^{N-1} x_j\ket{j} \mapsto \sum_{j=0}^{N-1} x_j\ket{j} \ket{f(j)},
\end{equation}
where $pixel$ and $channel$ (R, G or B) are classical parameters, $j$ plays the role of ray ID, and $f(j)$ is a real number that stands for the ray energy.
In the rest of this paper $f$ is specified as the function that maps ray ID to ray energy.
The oracle can trace $N=2^n$ paths simultaneously, and the final color we hope to write to the corresponding pixel and channel is the average of those energies,
\begin{equation}
    S = \frac{1}{N} \sum_{j=0}^{N-1} f(j).
    \label{eq:average}
\end{equation}

Suppose those real numbers $f(j)$ are stored in a fixed-point format with integer bit length $b_0$ and total bit length $b$, we can transfer the estimation problem of Eq.~\eqref{eq:average} into quantum counting by constructing a Boolean function,
\begin{equation}
    g(j, k) = \begin{cases}
        1,& f(j) \geq 2^{b_0-b}k; \\
        0,& f(j) < 2^{b_0-b}k.
    \end{cases}
    \quad
    (k=0,1,\cdots,2^b-1)
\end{equation}

The phase oracle $O_g$ for $g$ in Grover's search is,
\begin{equation}
    O_g: \sum_{j,k}\ket{j}\ket{k} \mapsto \sum_{j,k}(-1)^{g(j,k)}\ket{j}\ket{k}.
\end{equation}

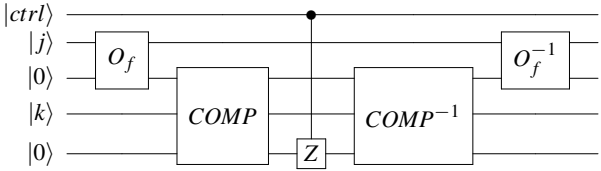
\begin{figure}[ht]
    $$\Qcircuit @C=1.2em @R=.6em {
        \lstick{\ket{ctrl}}     & \qw                & \qw                 & \ctrl{4} & \qw                      & \qw                     & \qw \\
        \lstick{\ket{j}} & \multigate{1}{O_f} & \qw                 & \qw      & \qw                      & \multigate{1}{O_f^{-1}} & \qw \\
        \lstick{\ket{0}} & \ghost{O_f}        & \multigate{2}{COMP} & \qw      & \multigate{2}{COMP^{-1}} & \ghost{O_f^{-1}}     & \qw \\
        \lstick{\ket{k}} & \qw                & \ghost{COMP}        & \qw      & \ghost{COMP^{-1}}        & \qw   & \qw \\
        \lstick{\ket{0}} & \qw                & \ghost{COMP}        & \gate{Z} & \ghost{COMP^{-1}}        & \qw   & \qw
    }$$
    \caption{ The construction of controlled-$O_g$, where the first qubit is the control qubit, and $Z:\ket{x}\mapsto(-1)^x\ket{x}$. }
    \label{fig:O_g}
\end{figure}

To construct $O_g$, we need a comparison gate that performs the comparison operation $COMP$ on the two integers $2^{b-b_0}f(j)$ and $k$,
\begin{equation}
    COMP: \sum_{j,k}\ket{f(j)}\ket{k} \mapsto \sum_{j,k}\ket{f(j)}\ket{k}\ket{g(j,k)},
\end{equation}
which is already constructed by \cite{QuantumAddition}.
With this in hand, the $O_g$ gate can be easily constructed, as shown in Fig.~\ref{fig:O_g}.

It is easy to verify that,
\begin{equation}
    S = \frac{1}{2^n}\sum_{j=0}^{N-1}f(j) = \frac{1}{2^{n+b-b_0}}\sum_{j=0}^{N-1}\sum_{k=0}^{2^b-1}g(j, k).
    \label{eq:sum-counting-relation}
\end{equation}

Also, the conversion to fixed-point format brings a truncation error of $O(2^{-b}N)$.

\begin{figure}[ht]
    $$\Qcircuit @C=1.5em @R=1.2em {
        \lstick{\ket{0}}   & \qw     & \ctrl{1}   & \qw                  & \ctrl{1}   & \qw                  & \qw \\
        \lstick{\ket{j,k}} & \qw{/^{n+b}} & \gate{O_g} & \gate{H^{\otimes (n+b)}} & \gate{O_0} & \gate{H^{\otimes (n+b)}} & \qw
    }$$
    \caption{ The construction of controlled-$G_g$ gate, where $\ket{j,k}$ is written as an entity of the two registers $\ket{j}$ and $\ket{k}$. }
    \label{fig:G_g}
\end{figure}
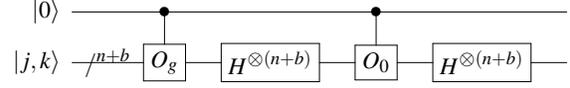

Finally, the quantity $\sum_{j,k}g(j, k)$ can be estimated by quantum counting algorithm.
The construction of controlled-$G_g$ required in quantum counting is shown in Fig.~\ref{fig:G_g}.
This is also where the dominant error of the whole procedure is brought.

%% \section{Performance}

\begin{figure*}
    \centering
    \includegraphics[width=\textwidth]{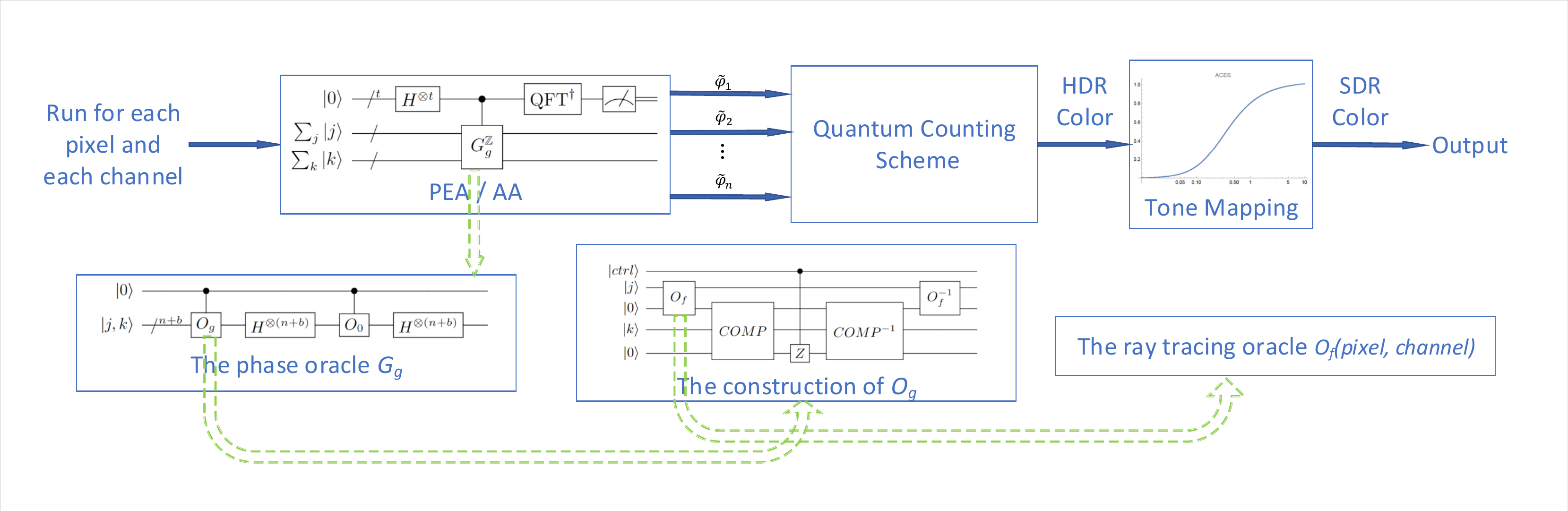}
    \caption{The framework of quantum ray tracing. The quantum counting may use PEA/AA for schemes. }
    \label{fig:framework}
\end{figure*}

In the paper we assume one call to $O_f$ in quantum realm takes the same time as tracing one path in classical realm.
We evaluate the cost of classical path tracing by the number of ray paths $N_c$, as the noise comes mostly from the Monte Carlo integration.
And in quantum ray tracing, the time cost is evaluated by the number of queries $N_q$ to the ray tracing oracle $O_f$, and the noise comes mostly from the random distribution of the output of PEA.
The QFT-based quantum counting has an error convergence rate of $O(1/N_q)$\cite{QuantumCounting1998}, hence has a quadratic speedup over classical Monte Carlo integration with convergence rate of $O(1/\sqrt{N_c})$.

% But the advantages of quantum ray tracing are more than that.
% The PEA noise in quantum ray tracing does not scale with the scene complexity, while the constant factor of error of classical Monte Carlo ray tracing does.
% Therefore, we believe that quantum ray tracing shows its supremacy in complex scenes.

Finally, the averaging step of ray energies takes place in the linear \textit{high-dynamic range} (HDR) color space.
To obtain a color between range $[0,1]$ in the \textit{standard-dynamic range} (SDR) space, a tone mapping\cite{ACES} step should be applied to the averaged color.
In summary, the framework of quantum ray tracing is illustrated in Fig.~\ref{fig:framework}.

\subsection{Schemes for quantum counting}

In many cases, QFT-based phase estimation algorithm (QFT-PEA) is already enough to use.
However, from Eq.~\eqref{eq:dist-PEA} we know $P(\tilde{\varphi}|\varphi)$ decays with an order of $\sin^{-2}(\pi(\tilde{\varphi}-\varphi))$ as $|\tilde{\varphi}-\varphi|$ grows big.
That is, though the output $\tilde{\varphi}$ of PEA is randomly distributed around the ground truth $\varphi$, it also has a long tail which shows as many detached noisy dots on the image.
That is exactly why the result of the original form of QSS contains many distinct noisy dots.

\begin{figure*}
    \centering
    \subfloat[Monte Carlo, $N_{\text{shot}}=1024$.]{
        \includegraphics[width=.3\textwidth]{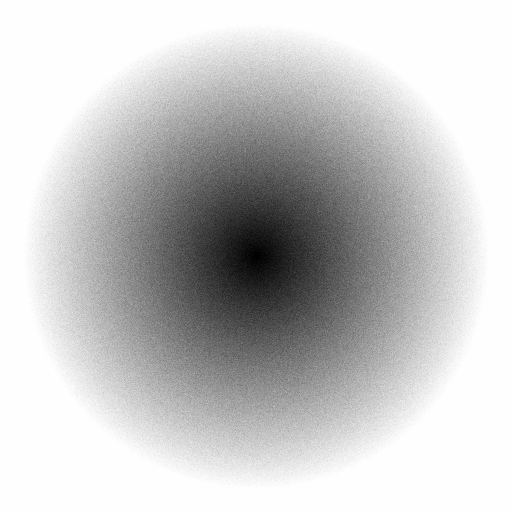}
        \label{fig:Gray_MC}
    }
    \subfloat[QFT-PEA, $T=1024$.]{
        \includegraphics[width=.3\textwidth]{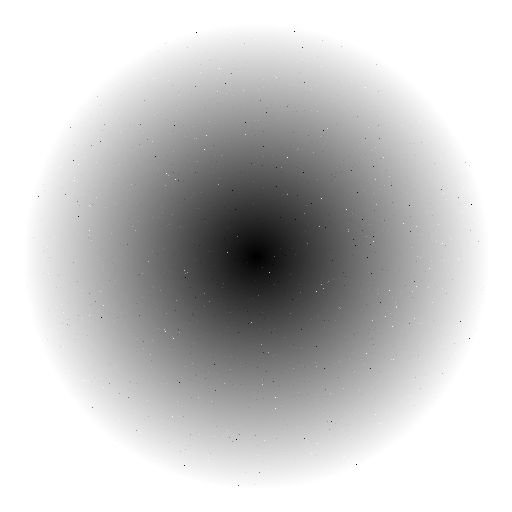}
        \label{fig:Gray_PEA}
    }
    \subfloat[QFT-BPEA, $T=256$, $N_{\text{shot}}=4$.]{
        \includegraphics[width=.3\textwidth]{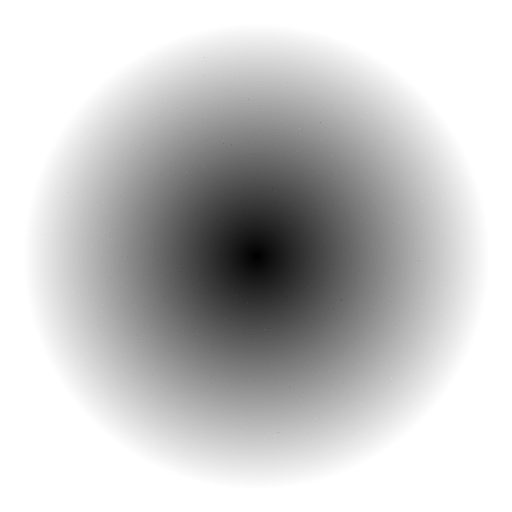}
        \label{fig:Gray_BPEA}
    }
    \caption{Gray disk experiments.}
    \label{fig:gray-disk}
\end{figure*}

To have a visual impression on this proposition, we do the experiments on the gray disk with the gray scale varying linearly from black at the center to white on the border, as shown in Fig.~\ref{fig:gray-disk}\ref{sub@fig:Gray_PEA}.
The gray scale of each pixel stands for $S/N$ in quantum counting.
By sampling from the theoretical random distribution of each algorithm, we can simulate the output $\tilde{S}/N$ and draw the gray scale to the corresponding pixel.
In Fig.~\ref{fig:gray-disk}, \ref{sub@fig:Gray_MC} shows the result of Monte Carlo sampling with $N$ samples drawn from the binary distribution as a comparison, \ref{sub@fig:Gray_PEA} shows the result of QFT-PEA, which contains some distinct dots, consistent with long tail phenomenon of Eq.~\eqref{eq:dist-QC}.

Intuitively, the noise of QFT-PEA can be reduced by repeating for $N_{\text{shot}}$ times and applying Bayesian estimation.
That is, finding the maximum of the function,
\begin{equation}
    L(\tilde\varphi) = \prod_{k=1}^{N_{\text{shot}}} P(\theta_k|\tilde\varphi),
    \label{eq:bpea}
\end{equation}
where we assume the priori distribution is uniform, $\{\theta_k\}$ are the estimated phases in each trial of PEA, and $P$ is the probability distribution given by Eq.~\eqref{eq:dist-QC}.
We call this method \textit{QFT-based Bayesian Phase Estimation} (QFT-BPEA) in this paper.
The gray disk experiment result is shown in Fig.~\ref{fig:gray-disk}\ref{sub@fig:Gray_BPEA}.
We can see that the noise level of detached distinct dots is obviously reduced.
But if we zoom in, Fig.~\ref{fig:gray-disk}\ref{sub@fig:Gray_BPEA} shows slight fake rings, because the parameter $T$ in each run of QFT-BPEA is smaller than QFT-PEA.
Thus, the fake ring phenomenon guided by Eq.~\eqref{eq:dist-PEA} in QFT-BPEA is stronger.

Based on the observation that about $8/\pi^2$ estimations of PEA are within accuracy $1/T$, we can dynamically adjust $N_{\text{shot}}$ until most of the samples are within an interval of length $2/T$.
We propose the \textit{QFT-based adaptive Bayesian PEA} (QFT-ABPEA) in Algorithm~\ref{alg:abpea}.

\begin{figure*}[ht]
    \centering
    \subfloat[Mean absolute error.]{
        \includegraphics[width=.45\textwidth]{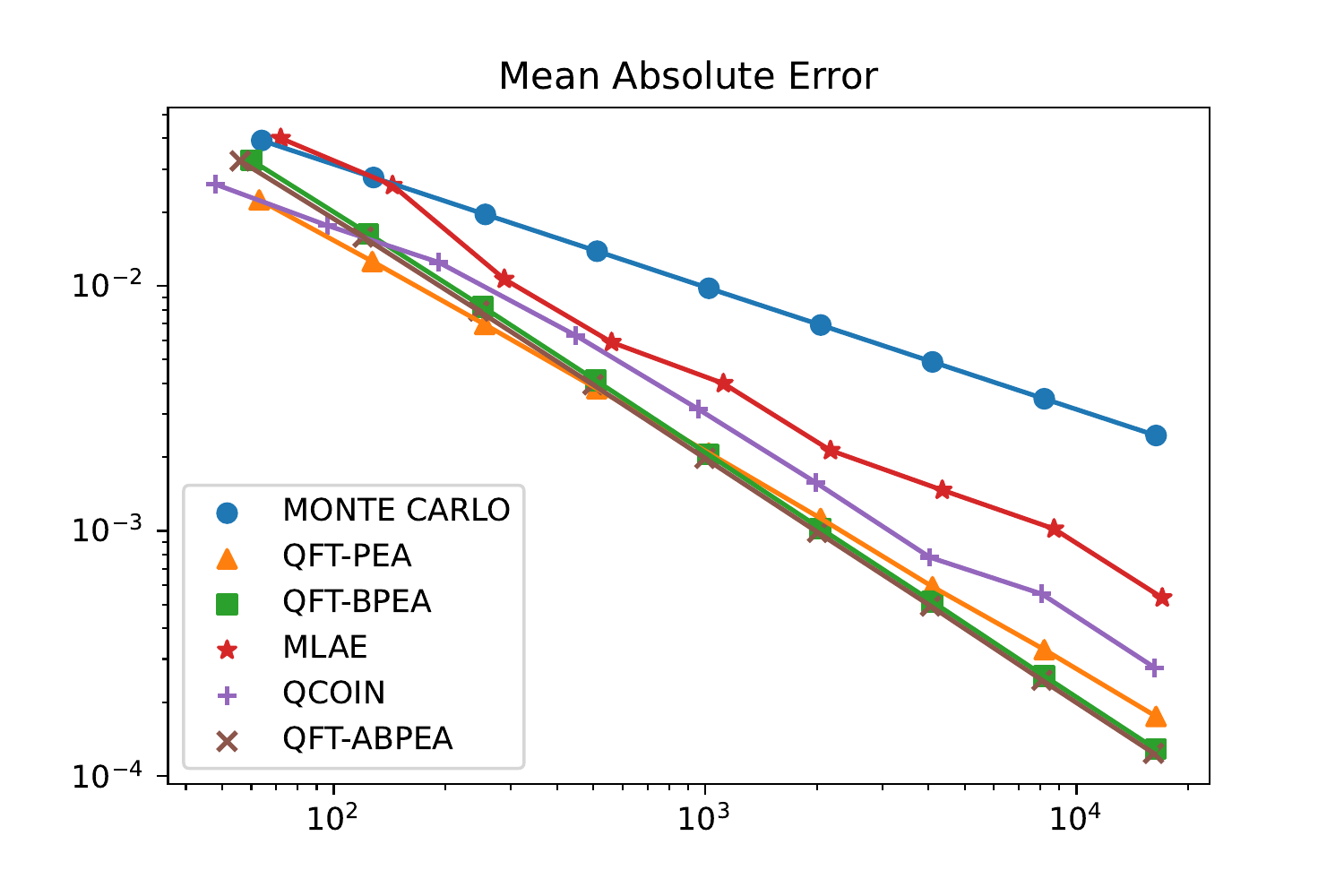}
        \label{fig:mean_absolute_error}
        \label{fig:rd_mae}
    }
    \quad
    \subfloat[Percentage of samples beyond the accuracy of 0.1.]{
        \includegraphics[width=.45\textwidth]{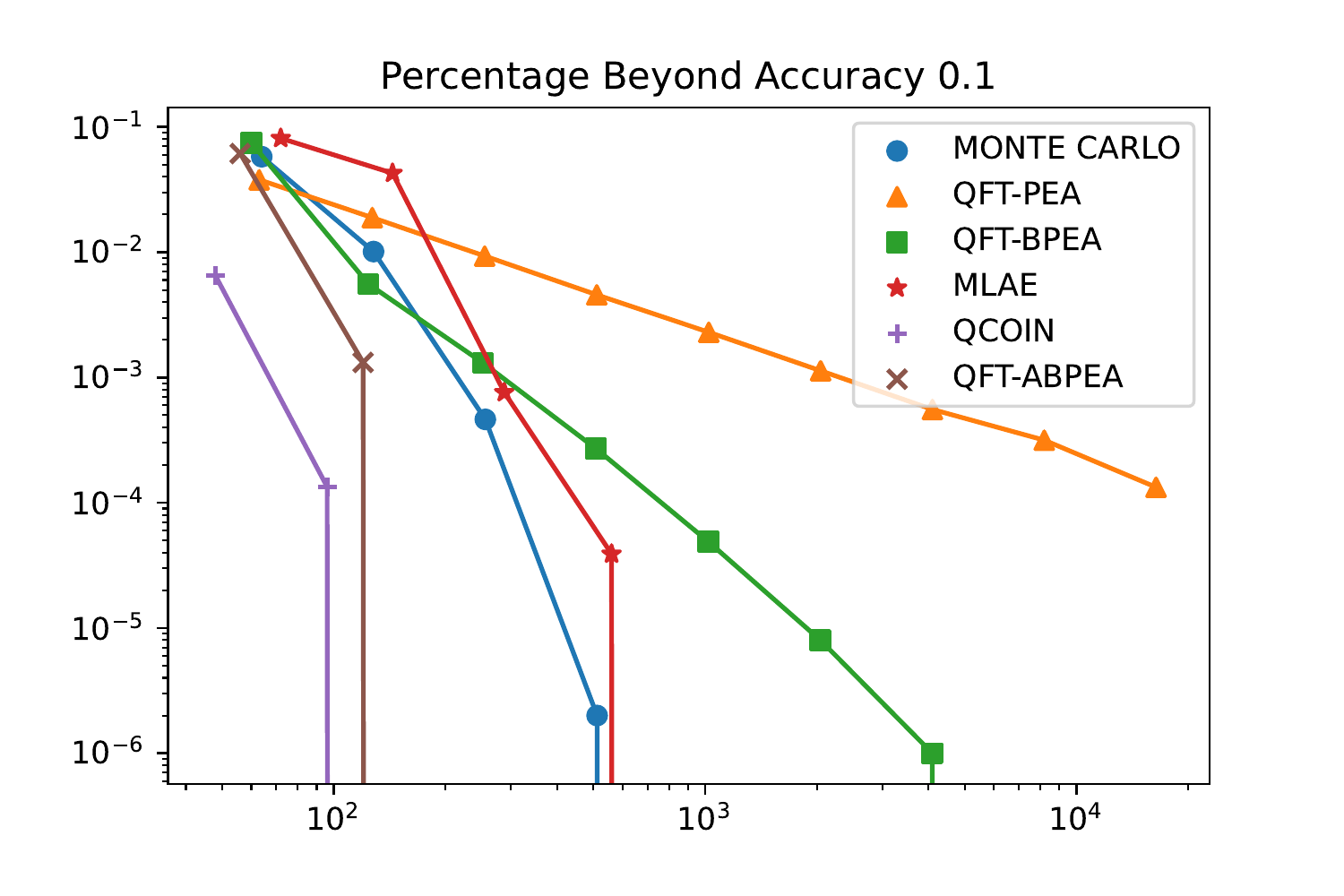}
        \label{fig:percentage_1}
        \label{fig:rd_pba1}
    }
    \quad
    \subfloat[Percentage of samples beyond the accuracy of 0.01.]{
        \includegraphics[width=.45\textwidth]{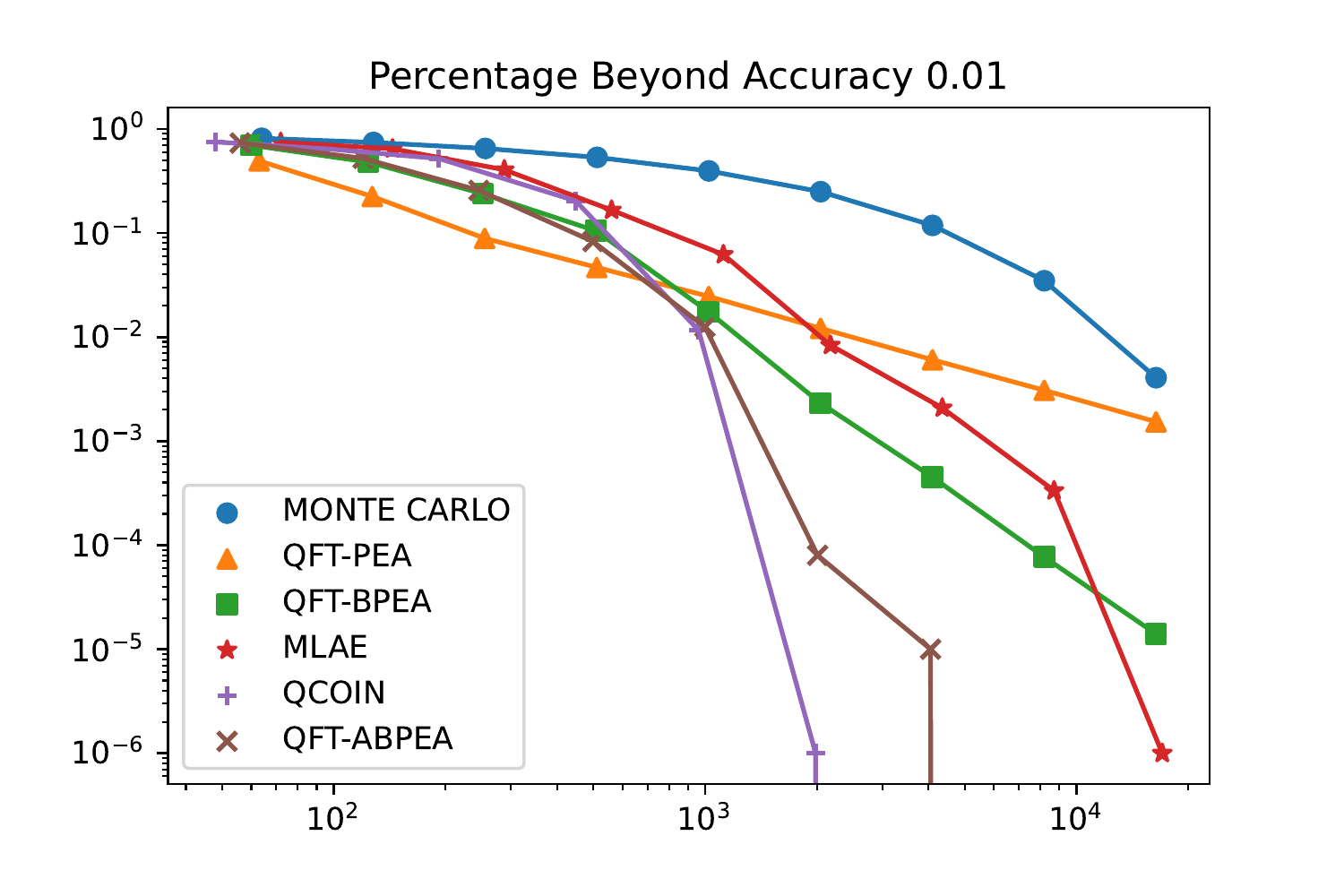}
        \label{fig:percentage_2}
        \label{fig:rd_pba2}
    }
    \quad
    \subfloat[Percentage of samples beyond the accuracy of 0.001.]{
        \includegraphics[width=.45\textwidth]{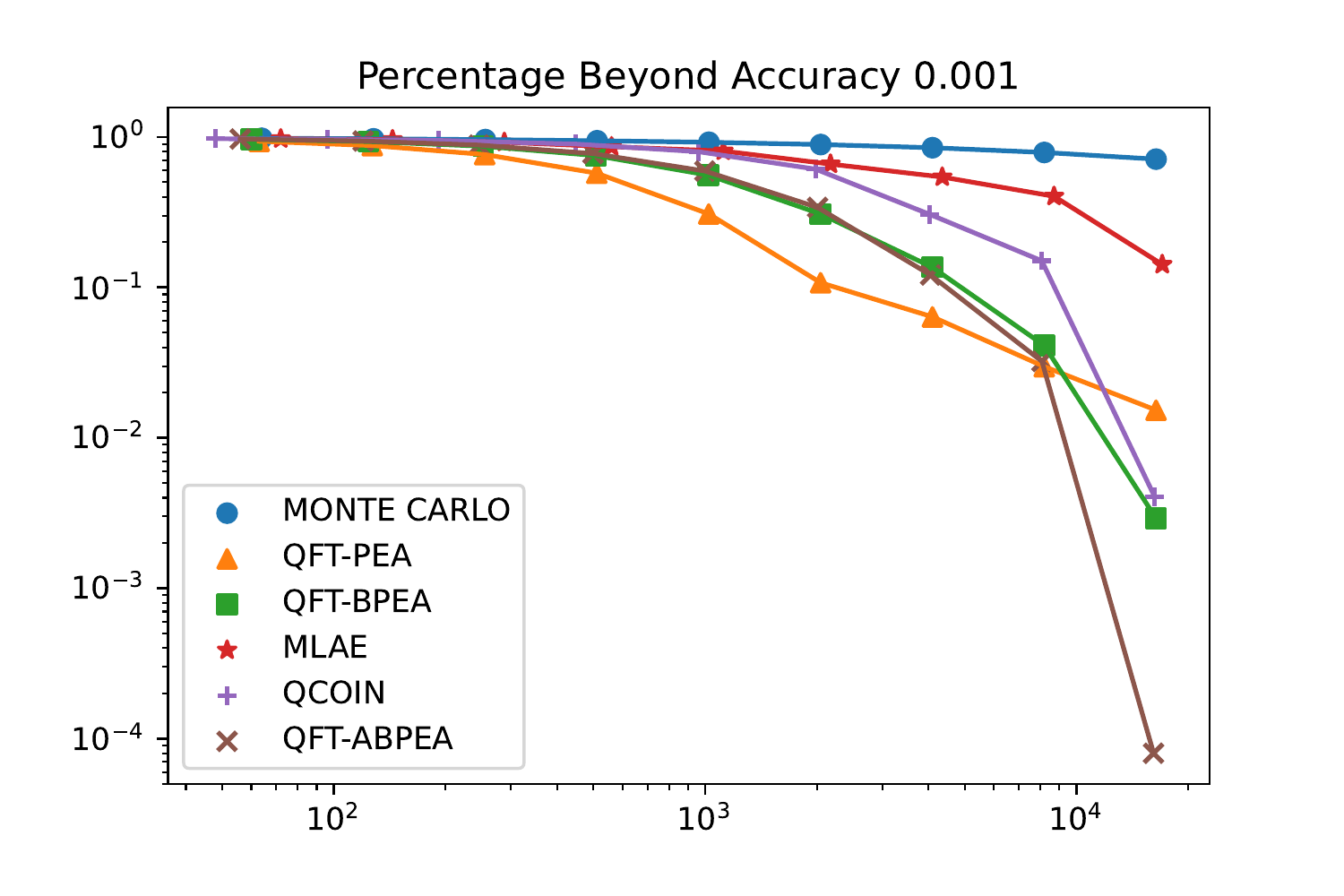}
        \label{fig:percentage_3}
        \label{fig:rd_pba3}
    }
    \caption{ Quantitative evaluations on how the error  scales with the number of queries to $G_g$ (the $x$-axis) for each scheme. }
    \label{fig:evaluation-query}
\end{figure*}

\begin{algorithm}
    \caption{QFT-ABPEA.}
    \label{alg:abpea}
    \KwIn{
        $\varphi$(visible in simulation experiments but invisible in real quantum computing), $T$,$\alpha$, $N_{\min}$, $N_{\max}$;
    }
    \KwOut{
        $\tilde\varphi$: the estimation of $\varphi$.
    }
    \BlankLine

    Initialize $\mathcal{S}=\{\tilde\varphi_1,\cdots\tilde\varphi_{N_{\min}}\}$ with $N_{\min}$ samples using QFT-PEA with parameter $T$;

    Sort $\mathcal{S}$ from small to big;

    \While{$\#\mathcal{S}<N_{\max}$ \textbf{and} \textnormal{There is no subset in $\mathcal{S}$ of length $\lfloor\alpha\#\mathcal{S}\rfloor$ and interval at most $2/T$}}{
        Insert one more sample into $\mathcal{S}$ while keeping $\mathcal{S}$ sorted;
    }

    Remove samples from $\mathcal{S}$ that is not in the interval;

    \Return{$\tilde\varphi$ = the Bayesian estimation from dataset $\mathcal{S}$};
\end{algorithm}

Besides the QFT family, we also introduce some other quantum counting schemes.

In 2019, Suzuki \etal~\cite{Suzuki2019} proposed an AA-based \textit{Maximum Likelihood Amplitude Estimation} (MLAE) algorithm.
They choose an exponentially incremental sequence of $M$, that is, choosing $M=2^m$ for $m\in\{0,1,\cdots,t-1\}$, estimating the probability of the amplitude with $M$ times of Grover's iteration for $N_{\text{shot}}$ times, and the final estimation is obtained by maximum likelihood estimation.
\begin{equation}
    \begin{aligned}
        L(\tilde\varphi) = \prod_{m=0}^{t-1} & \left[\cos^2\left((2^m+1)\pi\tilde\varphi\right)\right]^{N_{\text{shot}}-h_m}
        \\ \cdot & \left[\sin^2\left((2^m+1)\pi\tilde\varphi\right)\right]^{h_m}.
    \end{aligned}
\end{equation}
By finding the maximum of the likelihood function $L(\varphi)$ we obtain an estimation $\tilde\varphi$ of $\varphi$.

In 1999, Abrams \etal~\cite{Abrams1999} proposed an AA-based algorithm for quantum amplitude estimation, and the algorithm is later called \textit{quantum coin method} (QCoin) and numerical tested in \cite{Shimada2020}.
To estimate $\varphi\in[0,1]$ with high accuracy, they first estimate with Monte Carlo method to a smaller interval $[\varphi-\frac{\delta}{2}, \varphi+\frac{\delta}{2}]$, then remap the interval to $[0,1-\varepsilon]$ by amplitude amplification, and use Monte Carlo method to an even smaller interval.
By using different number $M$ of AA in each iteration, where $M=2^m$ for $m\in\{0,1,\cdots,t-1\}$, and each iteration using $N_{\text{shot}}$ repetitions, the information about $\varphi$ is extracted with faster error convergence than Monte Carlo method.
It should be mentioned that QCoin method uses a different framework from other schemes above.
QCoin method requires an additional step that transfers,
\begin{equation}
    \begin{aligned}
        & \sum_jx_j\ket{j}\ket{f(j)}\ket{0} 
        \\ \mapsto & \sum_jx_j\ket{j}\ket{f(j)}\left[\sqrt{1-\left(\frac{f(j)}{2^{b-b_0}}\right)^2}\ket{0}+\left(\frac{f(j)}{2^{b-b_0}}\right)\ket{1}\right],
    \end{aligned}
\end{equation}
whose accurate implementation requires $2^b$ extra controlled rotation gates.

To quantitatively analyze how noise scales with the number of queries in those schemes, we randomly generate $10^6$ numbers uniformly between 0 and 1 as ground truths of $S/N$, simulate the quantum counting algorithm with different schemes, and use the following statistical quantity to evaluate each scheme:
\begin{itemize}
    \item Mean Absolute Error: to evaluate the overall noise level.
    \item Percentage of samples beyond the accuracy of 0.1, 0.01 and 0.001: to evaluate the level of distinct noisy dots.
\end{itemize}

The results are shown in Fig.~\ref{fig:evaluation-query}.
In Fig.~\ref{fig:evaluation-query}\ref{sub@fig:rd_mae}, each quantum algorithm shows a faster error convergence than classical Monte Carlo algorithm.
In Fig.~\ref{fig:evaluation-query}\ref{sub@fig:rd_pba1}-\ref{sub@fig:rd_pba3} we evaluate the level of distinct noisy dots, where QCoin and QFT-ABPEA perform the best.

\begin{figure*}
    \centering
    \subfloat[Monte Carlo with $N_{\text{shot}}=1024$.]{
        \includegraphics[width=.3\textwidth]{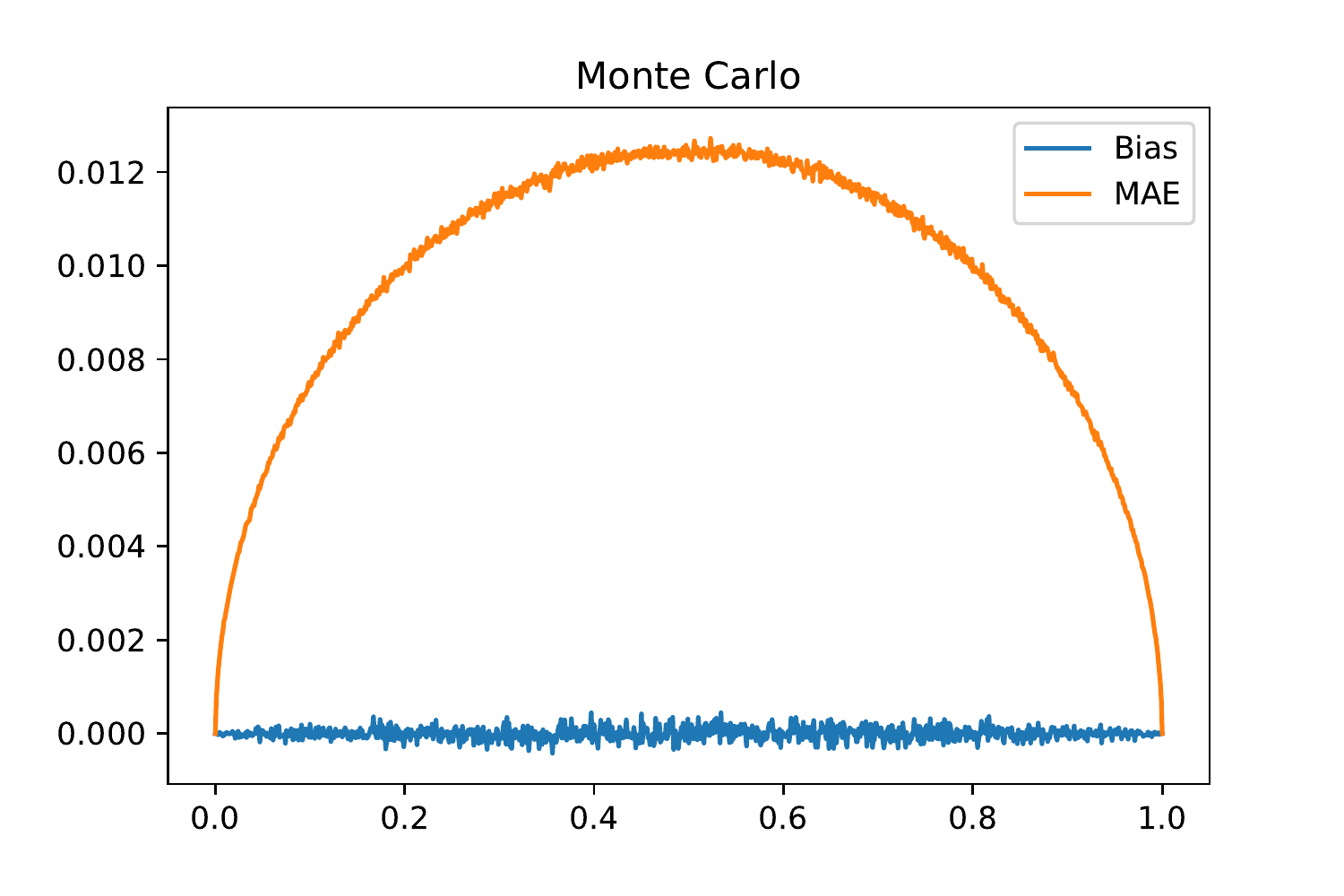}
        \label{fig:stat_mc}
    }
    \subfloat[MLAE with $T=16$ and $N_{\text{shot}}=64$.]{
        \includegraphics[width=.3\textwidth]{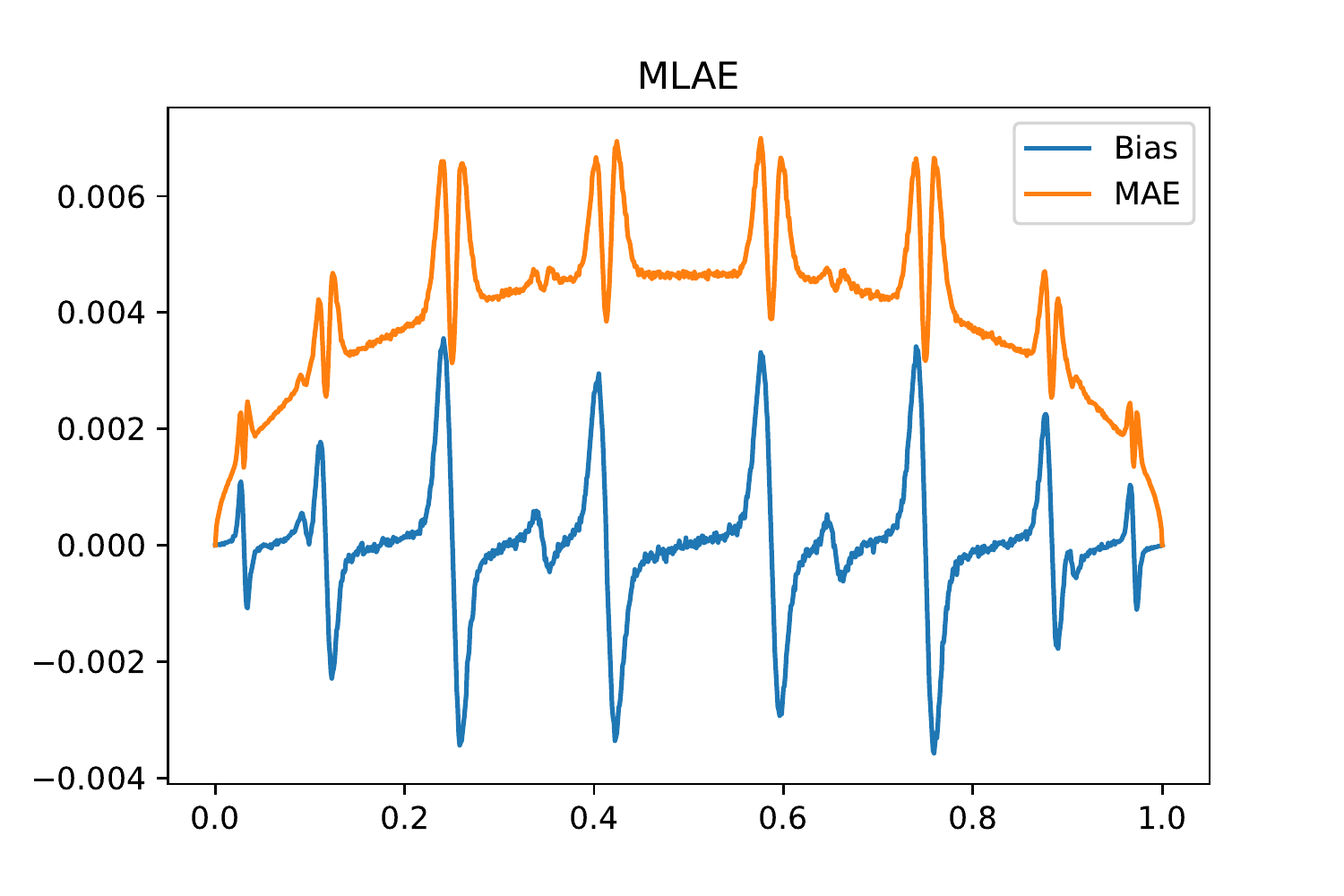}
        \label{fig:stat_mlae}
    }
    \subfloat[QCoin with $T=16$ and $N_{\text{shot}}=64$.]{
        \includegraphics[width=.3\textwidth]{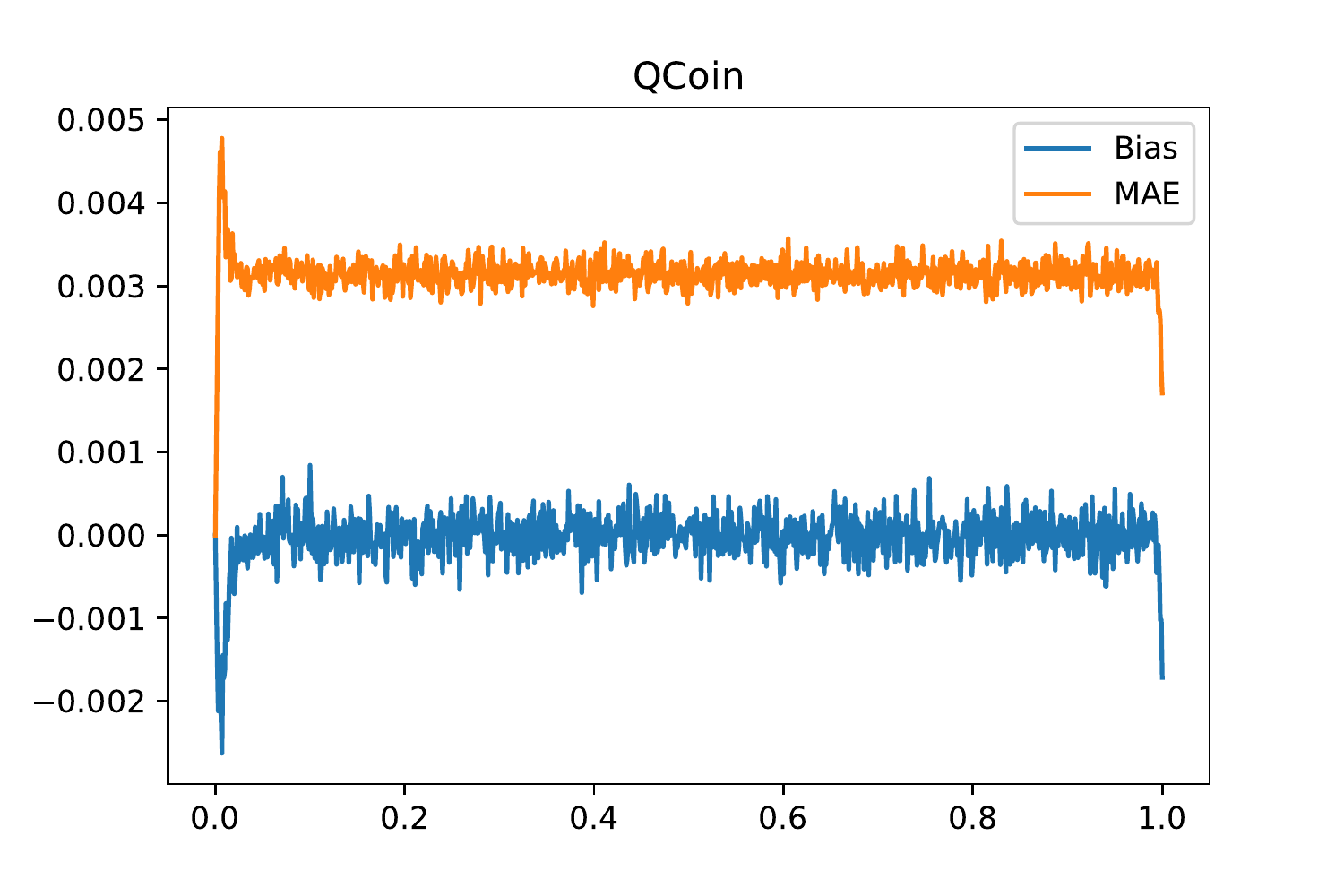}
        \label{fig:stat_qcoin}
    }
    
    \subfloat[QFT-PEA with $T=1024$.]{
        \includegraphics[width=.3\textwidth]{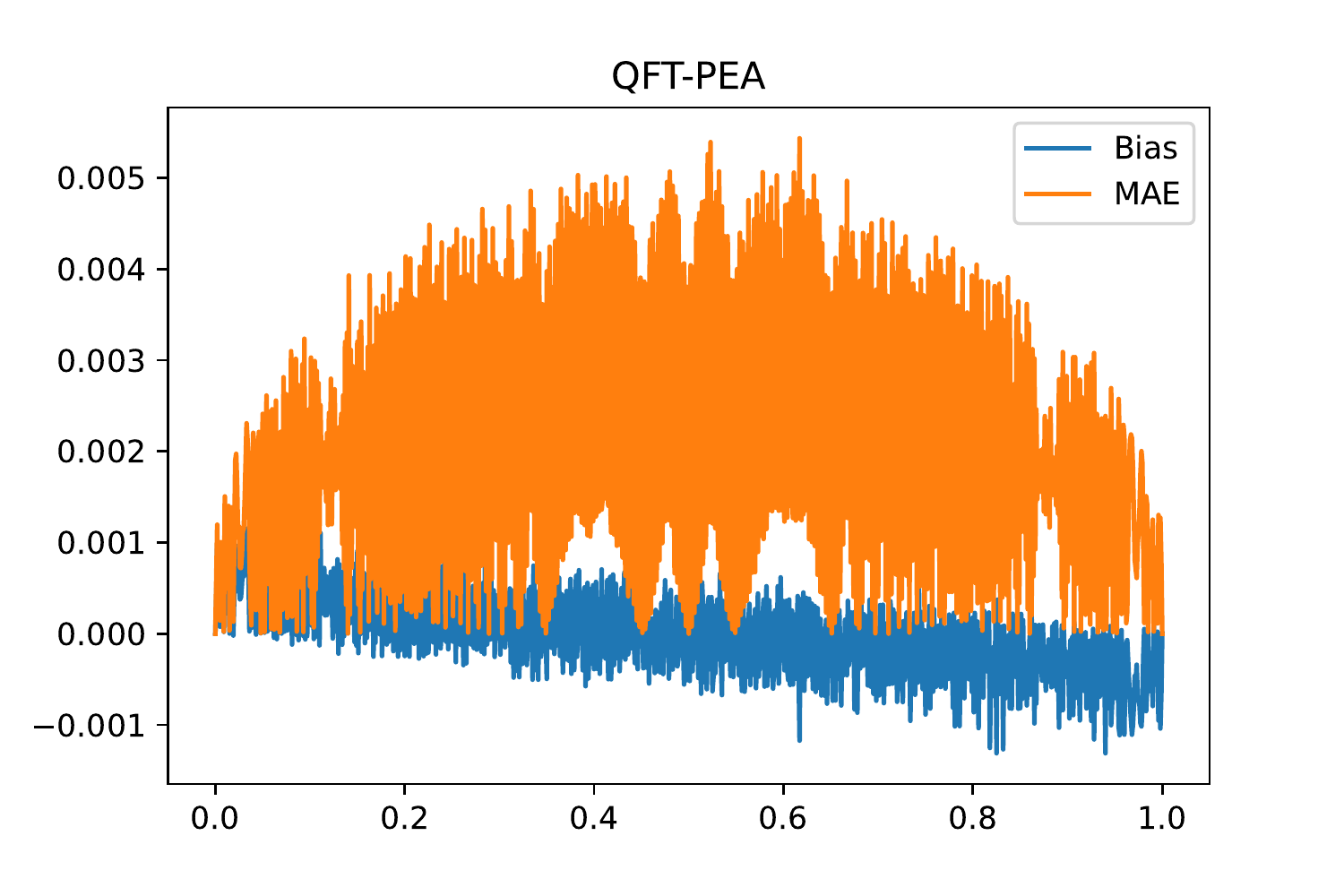}
        \label{fig:stat_pea}
    }
    \subfloat[QFT-BPEA with $T=256$ and $N_{\text{shot}}=4$.]{
        \includegraphics[width=.3\textwidth]{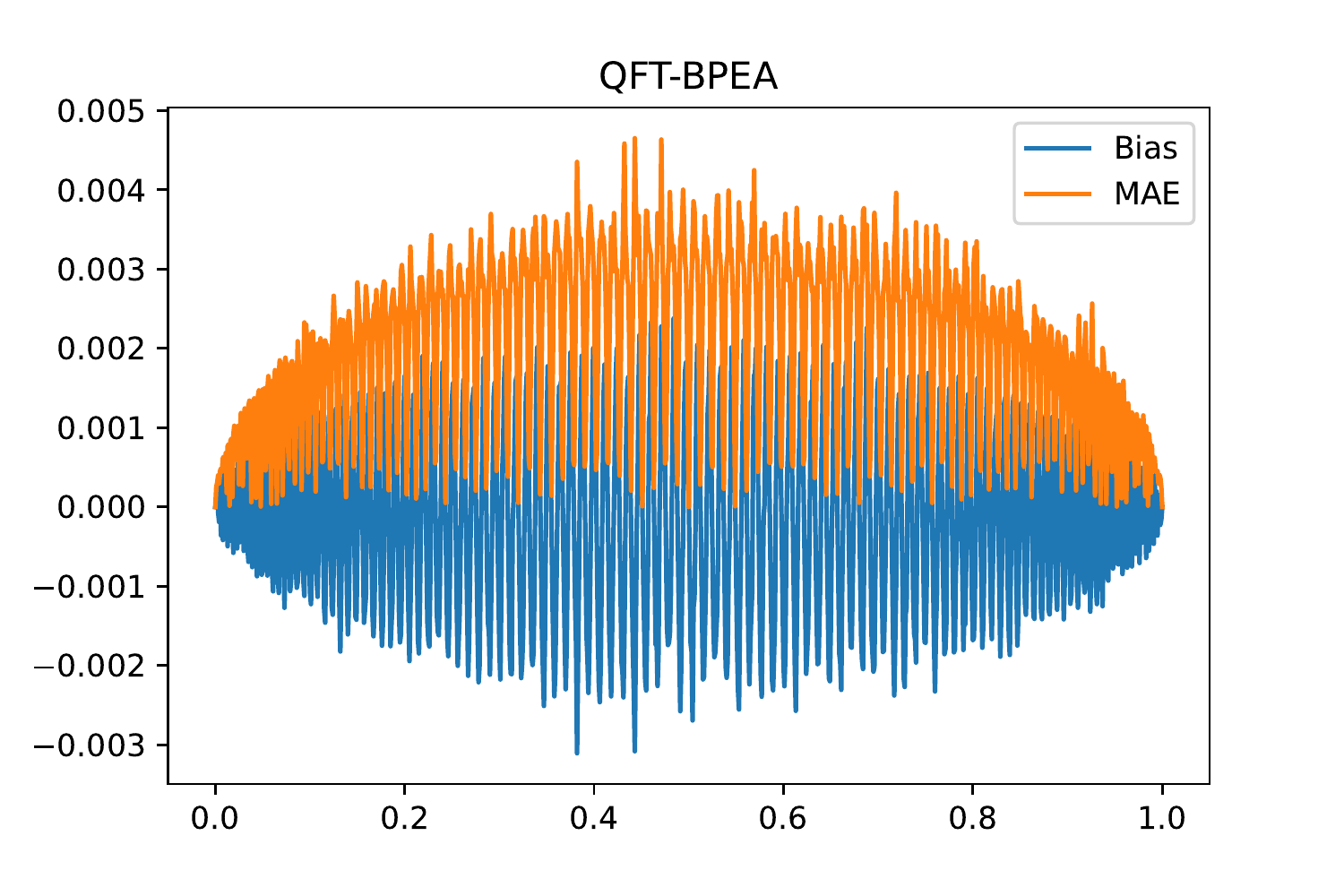}
        \label{fig:stat_bpea}
    }
    \subfloat[QFT-ABPEA with $T=256$, $N_{\min}=3$, $N_{\max}=8$, $\alpha=0.8$.]{
        \includegraphics[width=.3\textwidth]{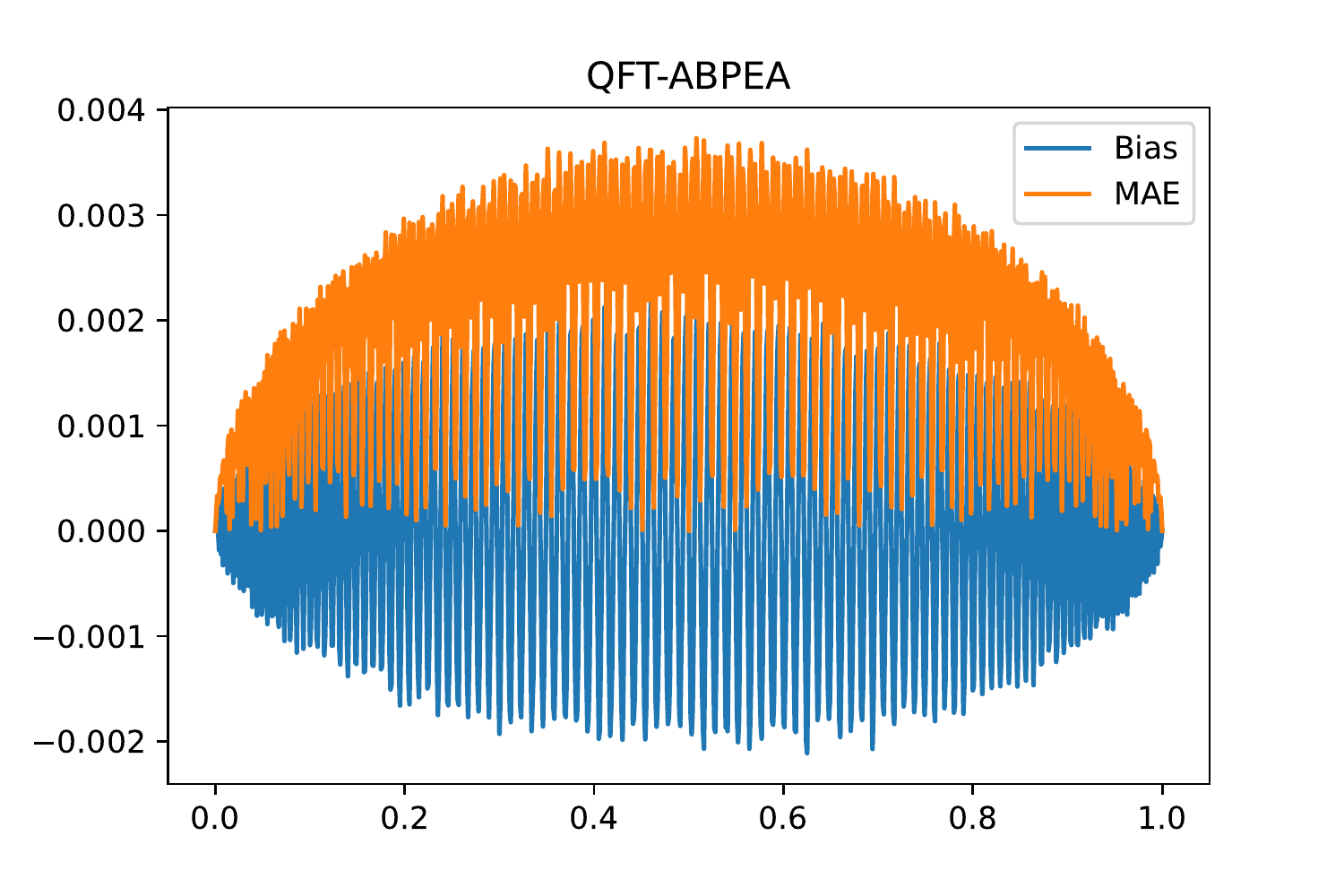}
        \label{fig:stat_abpea}
    }
    \caption{The error patterns of different schemes.}
    \label{fig:bias-mae}
\end{figure*}

Next, since the distribution of $\tilde\varphi$ of each quantum counting scheme is relevant to $\varphi$, we do another experiment to illustrate the error pattern of each scheme.
To be specific, for fixed $\varphi$s, we test each scheme for $N_{\text{test}}$ times, then analyze the bias
\begin{equation}
    \frac{\sum_{i=1}^{N_{\text{test}}}\tilde\varphi_i}{N_{\text{test}}} - \varphi,
\end{equation}
and the mean absolute error (MAE)
\begin{equation}
    \frac{1}{N_{\text{test}}}\sum_{i=1}^{N_{\text{test}}}\left|\tilde\varphi_i-\varphi\right|.
\end{equation}

We choose $\varphi\in\{0,0.001,0.002,\cdots,1\}$ and $N_{\text{test}}=10000$.
The results are shown in Fig.~\ref{fig:bias-mae}.
The Monte Carlo algorithm is the only unbiased one, but has the highest MAE.
The MLAE and QCoin have small bias in most area.
But MLAE has bias for some special values, which can also cause fake rings like QFT-based family.
And QCoin has bias and tends to give an underestimation around 0 or 1.
And both them have a moderate level of MAE.
The QFT-based family of algorithms have the smallest level of MAE, but the bias fluctuate greatly, which corresponds to the fake ring phenomenon.

\section{Experiment}
\label{sec:experiment}

In this section, we make comparisons on real images rendered by classical path tracing and simulated quantum ray tracing with different schemes.

\begin{figure*}
    \centering
    \includegraphics[width=.75\textwidth]{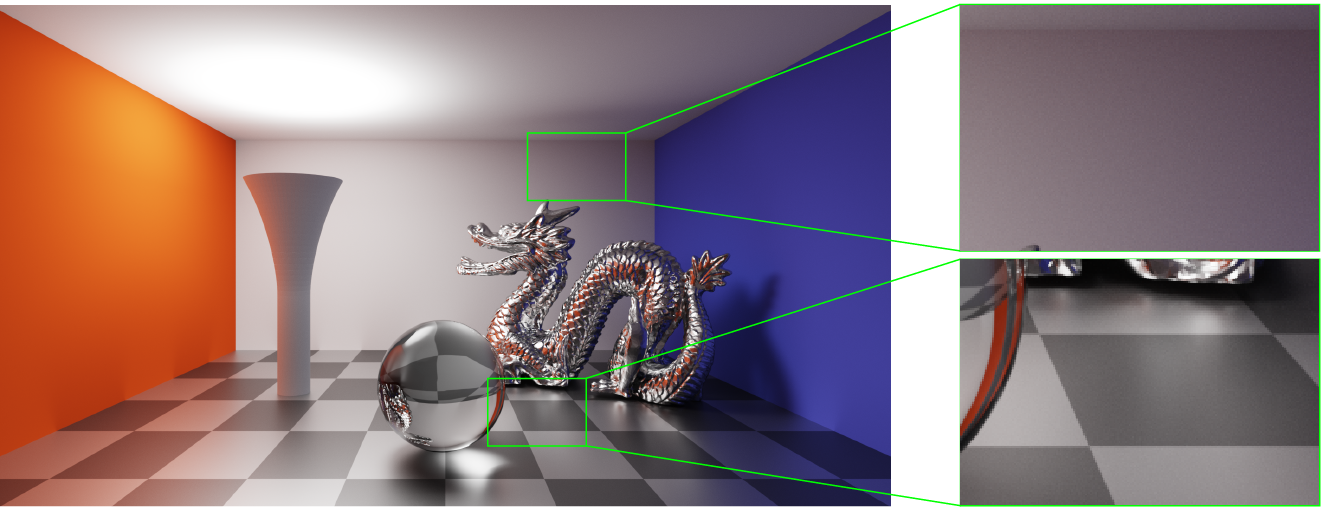}
    \caption{The ground truth, generated by Blender cycles with $2^{20}$ samples per pixel.}
    \label{fig:Dragon_GT}
\end{figure*}

\begin{figure*}
    \centering
    \includegraphics[width=.75\textwidth]{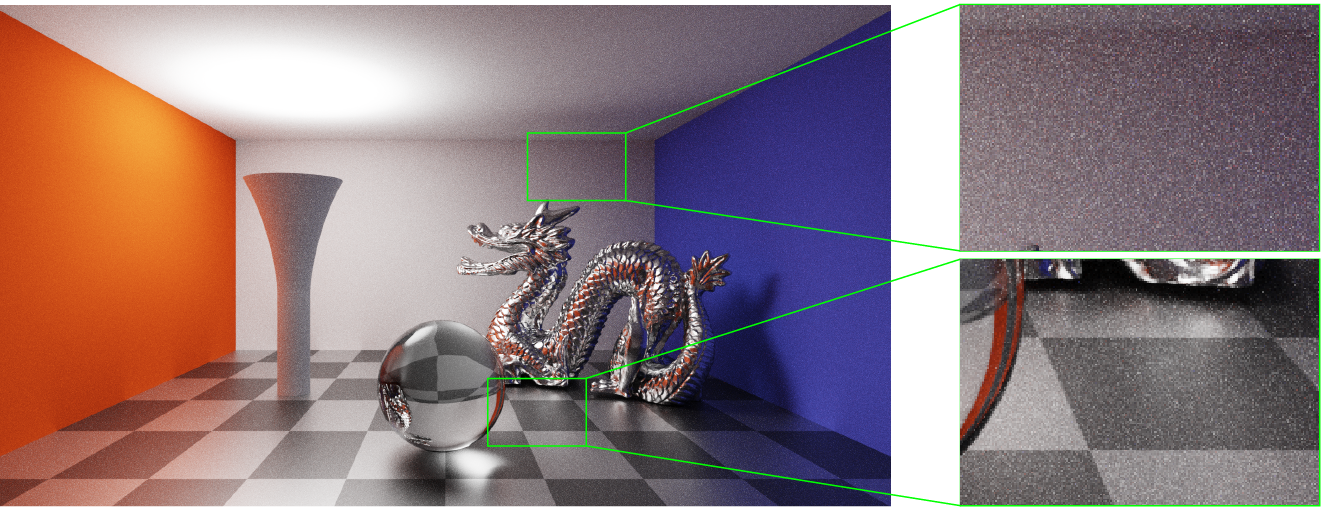}
    \caption{Classical ray tracing with 4250 sample per pixel. $N_c\approx 12186$.}
    \label{fig:Dragon_CL}
\end{figure*}

\begin{figure*}
    \centering
    \includegraphics[width=.75\textwidth]{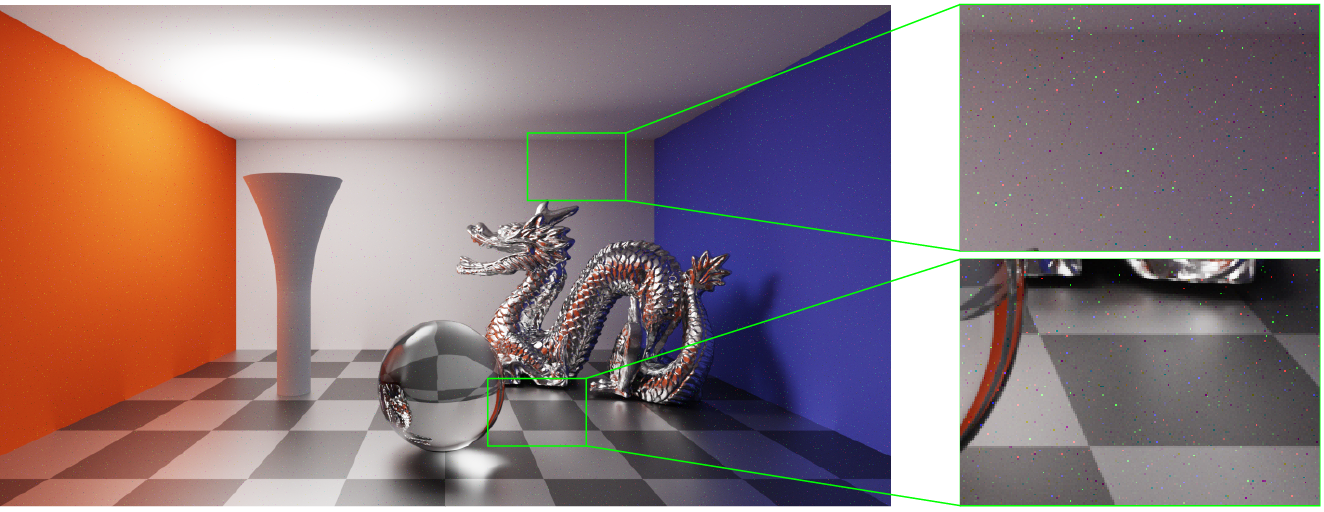}
    \caption{Quantum ray tracing result using QFT-PEA (QSS) with $T=2048$, $N_q=2047$.}
    \label{fig:Dragon_PEA}
\end{figure*}

\begin{figure*}
    \centering
    \includegraphics[width=.75\textwidth]{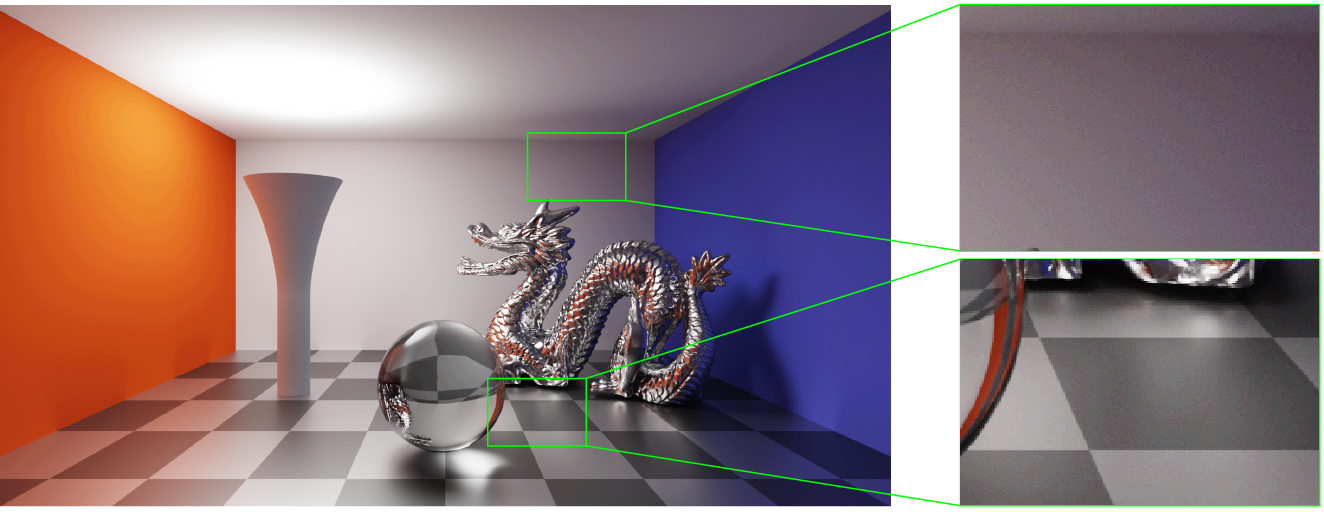}
    \caption{Quantum ray tracing using QFT-ABPEA with $T=512$, $\alpha=0.8$, $N_{\min}=3$ and $N_{\max}=8$, which reports $N_q=1983$.}
    \label{fig:Dragon_ABPEA}
\end{figure*}

\begin{figure*}
    \centering
    \includegraphics[width=.75\textwidth]{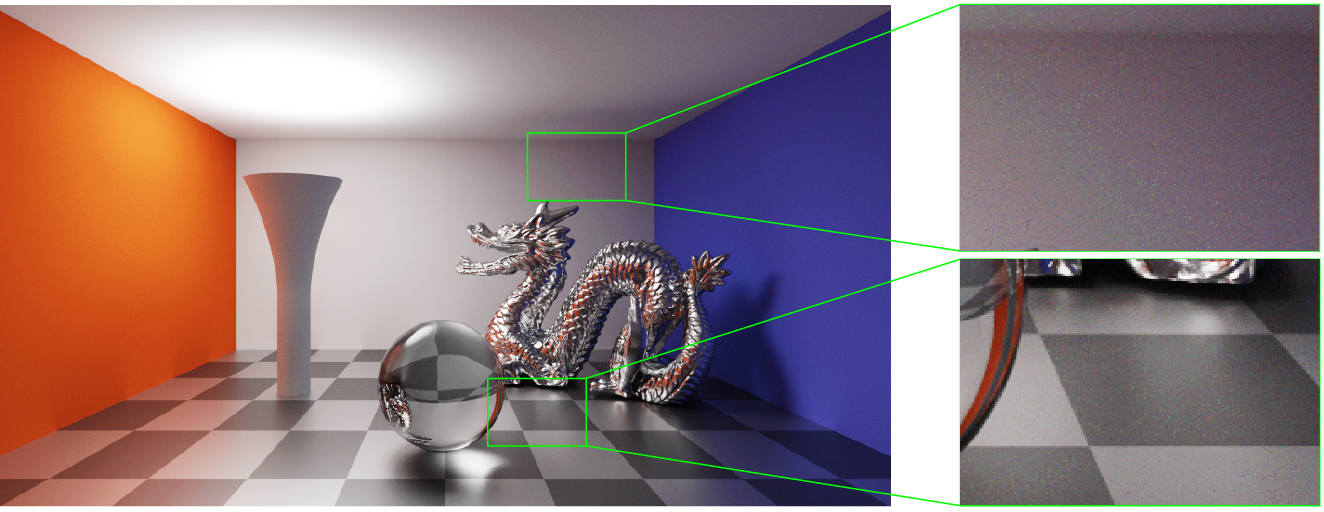}
    \caption{Quantum ray tracing using MLAE. $T=32$, $N_{\text{shot}}=64$. $N_q=2176$.}
    \label{fig:Dragon_MLAE}
\end{figure*}

\begin{figure*}
    \centering
    \includegraphics[width=.75\textwidth]{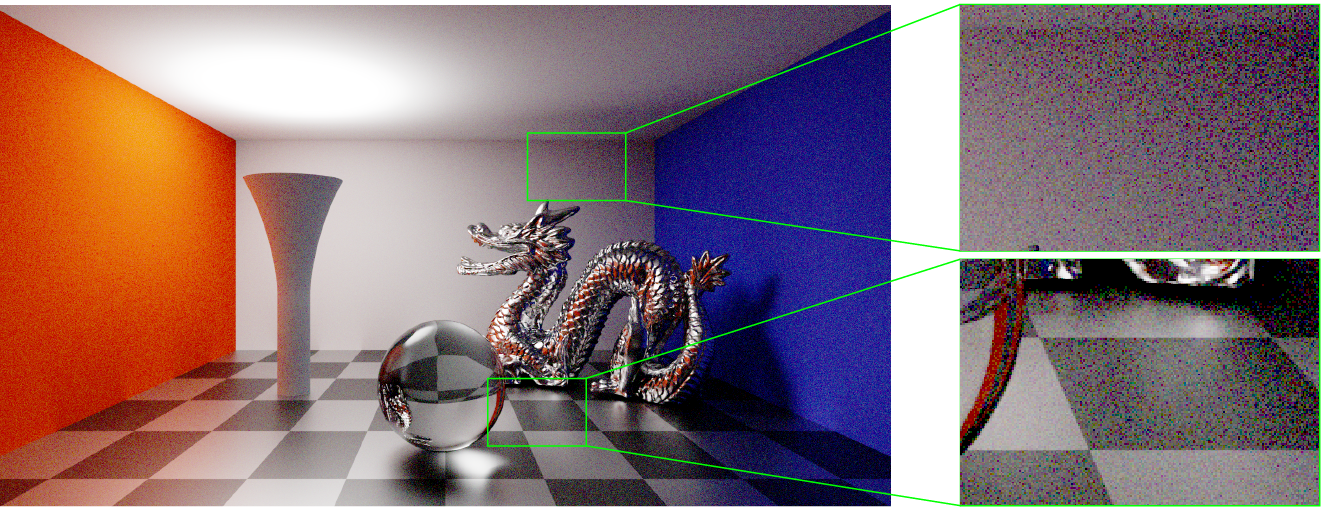}
    \caption{Quantum ray tracing using QCoin. $T=32$, $N_{\text{shot}}=64$. $N_q=2016$.}
    \label{fig:Dragon_QCoin}
\end{figure*}

We use the \textit{Blender cycles} renderer as a representation of the state-of-the-art classical ray tracing renderer.
As for the quantum ray tracing result, both actual quantum computers and classical simulators available now cannot provide enough quantum memory as well as coherence time for quantum ray tracing to show its power of superposing an astronomical number of rays.
Instead, we first render with a huge number of samples in Blender as the ground truth, output the colors in HDR space, and scale them into $[0,1]$.
Then we simulate the quantum noise with the same method in the pre-experiments in the scaled HDR space.
Finally, we scale them back, apply tone mapping and gamma correction to obtain the SDR color, and write these colors to the image.

We use the number of intersection searching sub-procedures as the measurement of the cost.
Though we can control the number of samples per pixel and the max tracing depth $D$, the actual number of intersection searching can only be computed by modifying the source code of Blender, since each ray may experience different tracing depth or direct light samples.
In comparison, the number of intersection searching of quantum ray tracing is completely definite, as all rays are traced as a superposition and thus share the same depth.
Suppose each $O_f$ traces the superposed rays to depth $D$, then one Grover's iteration $G_g$ contains $2D$ number of intersection searching, as one $G_g$ consists of one $O_f$ and one $O_f^{-1}$.
Suppose the quantum counting scheme requires $N_q$ queries to $G_g$, and notice that for each pixel there are three channels (RGB), then the calculation of the color of a single pixel requires $6DN_q$ number of intersection searching.

Denote $N_c$ to be the number of intersection searching divided by the number of pixels and $D$, as a measurement of the cost of classical ray tracing.
Then our experiment compares the rendering result of them conditioned on 
\begin{equation}
    N_c \approx 6N_q,
\end{equation}
as it is hard to force $N_c = 6N_q$.

We use \textit{Blender cycles} to render a scene with $2^{20}$ samples per pixel to depth $D=4$ as the ground truth, as illustrated in Fig.~\ref{fig:Dragon_GT}.
It should be mentioned that those ray IDs only require 20 extra qubits than storing a single ray in real quantum computers.
In this scene we use $b_0=4$ in Eq.~\eqref{eq:sum-counting-relation}, that is, each HDR color is scaled to 1/16 before simulating the quantum random distribution.
In addition, we use Blender cycles with $2^{20}$ samples per pixel as the ground truth, as a representation of classical ray tracing result, as shown in Fig.~\ref{fig:Dragon_CL}.
We simulate the quantum ray tracing result, with QFT-PEA in Fig.~\ref{fig:Dragon_PEA}, QFT-ABPEA in Fig.~\ref{fig:Dragon_ABPEA}, MLAE in Fig~\ref{fig:Dragon_MLAE} and QCoin in Fig.~\ref{fig:Dragon_QCoin}.

In this experiment, while the classical ray tracing result is still noisy, the quantum ray tracing results can be already clean enough.
The QFT-PEA result with $T=2048$ shows many distinct noisy dots, but shows good smoothness in other area.
The QFT-ABPEA result contains much slighter distinct dots, but shows slight fake rainbow-like strips.
The MLAE result shows a moderate level of fake rings and smoothness.
Finally, the result of QCoin shows obvious error, since the HDR color of most pixels are not exposed and thus becomes close to zero when divided by $2^{b_0}$, where QCoin does not behave well as illustrated in Fig.~\ref{fig:bias-mae}.

\section{Discussion}
\label{sec:discussion}

Here we discuss the limitations and potential improvements of the application of improved QSS in quantum ray tracing.

First, we claim that we are comparing quantum ray tracing and classical ray tracing by comparing the visual noise level conditioned on similar numbers of queries.
In other words, we assume that the costs of performing an intersection searching in both quantum ray tracing and classical ray tracing are the same.
However, classical ray tracing uses data structures like BSP tree~\cite{Fuchs1980}, KD-tree~\cite{Bentley1975, hapala2011kd} and BVH~\cite{Clark1976} to avoid traversing all scene primitives.
Meanwhile, in quantum ray tracing where an astronomical number of rays are superposed, we have to traverse all scene primitives.
Besides, we do not take GPU parallel computing into consideration, which is another vital acceleration technique in classical ray tracing.
In short, we may overestimate the time cost of classical ray tracing, and thus overestimate the speedup of quantum ray tracing.

Second, since the rendering equation is infinitely recursive and any cutoff on the ray tree will bring numerical truncation error, classical ray tracing utilizes Russian Roulette to ensure the unbiasedness, but in quantum ray tracing all superposed rays share the same ray tree depth, and we have no access to the intermediate information to decide the halt condition dynamically and have to pre-determine a fixed ray tree, which may cause visual artifacts and must be compensated for.
But in the experiments of this paper we do not simulate anything about that.

Third, quantum ray tracing only shows its power when the scene complexity is large enough.
As we have addressed, the noise level of quantum ray tracing is irrelevant to the scene complexity.
If a moderate level of ray samples in classical ray tracing can reduce the noise to an acceptable level, then the quantum ray tracing result with the same number of queries just brings extra noise.
In real-time ray tracing where computational resources are strictly limited, quantum ray tracing may not perform better.

There are also potential improvements in our quantum ray tracing algorithm.
In the current framework we only superpose the rays, and the information is extracted by phase estimation.
Intuitively, the scene primitives can be another potential candidate for superposition.
By leveraging minimum finding algorithm~\cite{Durr1996}, the information of the intersection with the minimum distance can be extracted, with another quadratic speedup as well as additional error probability.
However, the two frameworks work alone and cannot be directly joined, since minimum finding algorithm works in a hybrid quantum-classical way and should be implemented for each ray, where the advantage of superposing an astronomical number of rays in the current framework will fail.
It will be another interesting topic to join the two frameworks to obtain a biquadratic speedup.

\section{Conclusion}
\label{sec:conclusion}

In the original form of QSS~\cite{Eric2016}, the image produced by quantum supersampling contains many detached noisy dots, as illustrated in Fig.~\ref{fig:Dragon_PEA}.
In this paper, we improve quantum supersampling for quantum ray tracing, by replacing the standard QFT-based phase with more robust quantum counting schemes like QFT-BPEA, QFT-ABPEA and MLAE.
We also take another quantum counting scheme, QCoin, into comparison.
We quantitatively evaluate those schemes with visual performance related evaluations, and study how they scale with the parameters of the schemes as well as their error patterns.

Finally, we build a 3D scene, use $2^{20}$ samples per pixel with Blender cycles renderer to generate the ground truth, and then simulate the quantum noise in the HDR linear space by sampling random numbers with respect to the theoretical distribution, to simulate the quantum ray tracing result.
We show that by choosing appropriate schemes the image quality can be much better than classical ray tracing and original QSS.

\bibliographystyle{eg-alpha-doi}
\bibliography{ref}

\end{document}